\documentclass[usenatbib]{mn2e}

\pdfoutput=1

\usepackage[pdftex]{graphicx} 

\usepackage{times}

\usepackage{rotate} 

\usepackage{amssymb} 

\usepackage{rotating}

\usepackage{epstopdf} 

\usepackage{graphicx, subfigure}

\usepackage{array}

\title[Obscuration properties of mid-IR selected AGN]{Obscuration properties of mid-IR selected AGN}

\author[G. Mountrichas, I. Georgantopoulos, A. Ruiz, G. Kampylis ]{G. Mountrichas$^1$, I. Georgantopoulos$^1$, A. Ruiz$^1$, G. Kampylis$^{1,2}$ \\ \\ 
$^1$National Observatory of Athens, V.  Paulou  \& I.  Metaxa, 15236,  Greece \\
$^2$Department of Physics, University of Patras, Patras, Greece}

\begin{document}
\maketitle
\label{firstpage}

\begin{abstract}  
The goal of this work is to study the obscuration properties of mid-infrared (mid-IR) selected AGN. For that purpose, we use {\it{WISE}} sources in the Stripe 82-XMM area to identify mid-IR AGN candidates, applying the Assef et al. criteria. Stripe 82 has optical photometry $\approx$\,2 times deeper than any single-epoch SDSS region. XMM-Newton observations cover $\sim$26\,deg$^2$. Applying the aforementioned criteria, 1946 IR AGN are selected. $\sim 78\%$ have SDSS detection, while 1/3 of them is detected in X-rays, at a flux limit of $\rm \sim 5 \times 10^{-15}\,erg\,s^{-1}\,cm^{-2}$. Our final sample consists of 507 IR AGN with X-ray detection and optical spectra. Applying a $\rm r-W2>6$ colour criterion, we find that the fraction of optically red AGN drops from 43\% for those sources with SDSS detection to $23\%$ for sources that also have X-ray detection. X-ray spectral fitting reveals 40 ($\sim8\%$) X-ray absorbed AGN ($\rm N_H>10^{22}~cm^{-2}$). Among the X-ray unabsorbed AGN, there are 70 red systems. To further investigate the absorption of these sources, we construct Spectral Energy Distributions (SEDs) for the total IR AGN sample. SED fitting reveals that $\sim20\%$ of the optically red sources have such colours because the galaxy emission is a primary component in the optical part of the SED, even though the AGN emission is not absorbed at these wavelengths. SED fitting also confirms that $12\%$ of the X-ray unabsorbed, IR AGN are optically obscured. 
\end{abstract}


\section{Introduction}

Mid-infrared (mid-IR) surveys have been proven extremely capable of detecting Active Galactic Nuclei (AGN), since they are affected less by extinction. The material that obscures AGN even at X-ray energies, is heated by the AGN and re-emits the nuclear radiation at infrared (IR) wavelengths. Thus, mid-IR surveys trace obscured sources missed even by hard X-ray surveys \citep[e.g.,][] {Georgantopoulos2008, Fiore2009}. {\it{Spitzer}} was the first IR mission used to demonstrate the efficiency of selecting AGN from a mid-IR dataset, by applying colour-selection techniques \citep[e.g.,][]{Stern2005, Donley2012}. These techniques have now been adapted and used for the {\it{Wise-field Infrared Survey Explorer}} \citep[{\it{WISE}};][]{Wright2010}.

{\it{WISE}} completed an all-sky coverage in four mid-IR bands 3.4, 4.6, 12 and 12\,$\mu$m (W1, W2, W3 and W4 bands, respectively). Several colour regions have been defined that efficiently identify AGN, in particular at high luminosities. For example, \cite{Mateos2012} suggested a selection method using three {\it{WISE}} colours. \cite{Stern2012} used the W1 and W2 bands and applied a simple $\rm{W1-W2}\geq 0.8$ criterion to reliably, select AGN with $\rm{W2}<15.05$ in the COSMOS field. \cite{Assef2013} extended the aforementioned criterion and provided a selection of AGN for fainter {\it{WISE}} sources. 

Previous studies have shown that a fraction of these IR selected AGN are missed by X-rays. This is attributed to their heavy obscuration in the X-ray band. \cite{delmoro2016} studied a sample of 33 mid-IR selected quasars with intrinsic luminosity $\nu\,L_{6 \mu m} > 6 \times 10^{44}$\,erg\,s$^{-1}$ at redshift $z\approx 1-3$. Despite their high IR luminosity, $\sim 30\%$ of the sources, i.e. nine quasars, are undetected in X-rays. Among the X-ray detected IR sources, 16 out of 24 ($\sim70\%$) are heavily obscured (N$_H>2\times 10^{23}$\,\rm{cm}$^{-2}$). \cite{Mateos2017} used 199 X-ray selected AGN from the Bright Ultra-hard XMM-Newton Survey (BUXS) and found a substantial population of X-ray undetected sources with high-covering factor tori. They claim that the majority of luminous AGN live in highly obscured environments that remain undetected in X-rays at the depths of $<10$\,keV wide-area surveys. \cite{Mendez2013} found that the percentage of IR AGN that are also detected in X-rays, varies significantly (47\% to 90\%) depending on the depth of the X-ray and IR surveys (see their Fig. 8). Increasing the depth of the IR data reduces the fraction of X-ray detected sources, while increasing the depth of the X-ray data increases the fraction of X-ray detections.

\cite{Secrest2015} applied the \cite{Mateos2012} colour selection criteria on the {\it{WISE}} sample and created a sample of 1.4 million AGN candidates. \cite{Mountrichas2017a} used this catalogue to explore the X-ray properties of mid-IR selected AGN. In this work they study only the most luminous of these  sources ($\rm{log}\,(\nu L_\nu /erg\,s^{-1})\ge 46.2$). Specifically, they cross-correlated the \cite{Secrest2015} catalogue with the subsample of the 3XMM X-ray dataset with available X-ray spectra \citep{Corral2015} and optical spectroscopy from SDSS/BOSS \citep{Alam2015}. Due to the requirement for optical (SDSS) identifications, their sample is  biased towards type-1 sources. Their analysis revealed seven obscured AGN in X-rays. However, none of them was absorbed based on their optical continuum. 

\cite{Lamassa2019} studied the demographics of AGN in Stripe82X \citep{LaMassa2013a, LaMassa2013b}. They compiled a catalogue of 4847 AGN, based on their X-ray luminosities or {\it{WISE}} colours. Stripe 82X has dedicated SDSS observations, therefore their sample does not suffer from the optical spectroscopic limitations of the  \cite{Mountrichas2017a} dataset. Their analysis showed that 61\% of X-ray AGN are not selected as mid-IR AGN, while 22\% of X-ray sources have no {\it{WISE}} detection. Moreover,  58\% of {\it{WISE}} AGN are not detected in X-rays. However, based on the W$1-$W2 colour difference, sources undetected in X-rays do not appear to be redder than those detected in X-rays. Previous studies have shown, that although X-ray selection is able to identify even very inactive AGN, the X-ray emission is easily absorbed by gas. Especially, at the softer, $10$\,KeV energies probed by the Chandra and XMM satellites. On the other hand, mid-IR selection is less susceptible to absorption, but only identifies high luminosity AGN, i.e., those systems that the AGN outshines the stellar emission from the galaxy \citep[e.g.,][]{Barmby2006, Georgantopoulos2008, Eckart2010}.

  
In this work, we use {\it{WISE}} sources in the Stripe 82X survey \citep{LaMassa2015,Lamassa2019} and apply the colour criteria of \cite{Stern2012}, modified by \cite{Assef2013} to select mid-IR AGN candidates. First, we  derive the obscuration properties using the optical/mid-IR colours and Spectral Energy Distributions (SED). Second, we use only those sources with X-ray detection. Our goal is to study the X-ray properties of these mid-IR selected AGN, by fitting their X-ray spectra. We also compare the X-ray with the optical colour of these sources.

\begin{figure*}
\centering
\begin{subfigure}
  \centering
  \includegraphics[width=0.32\linewidth]{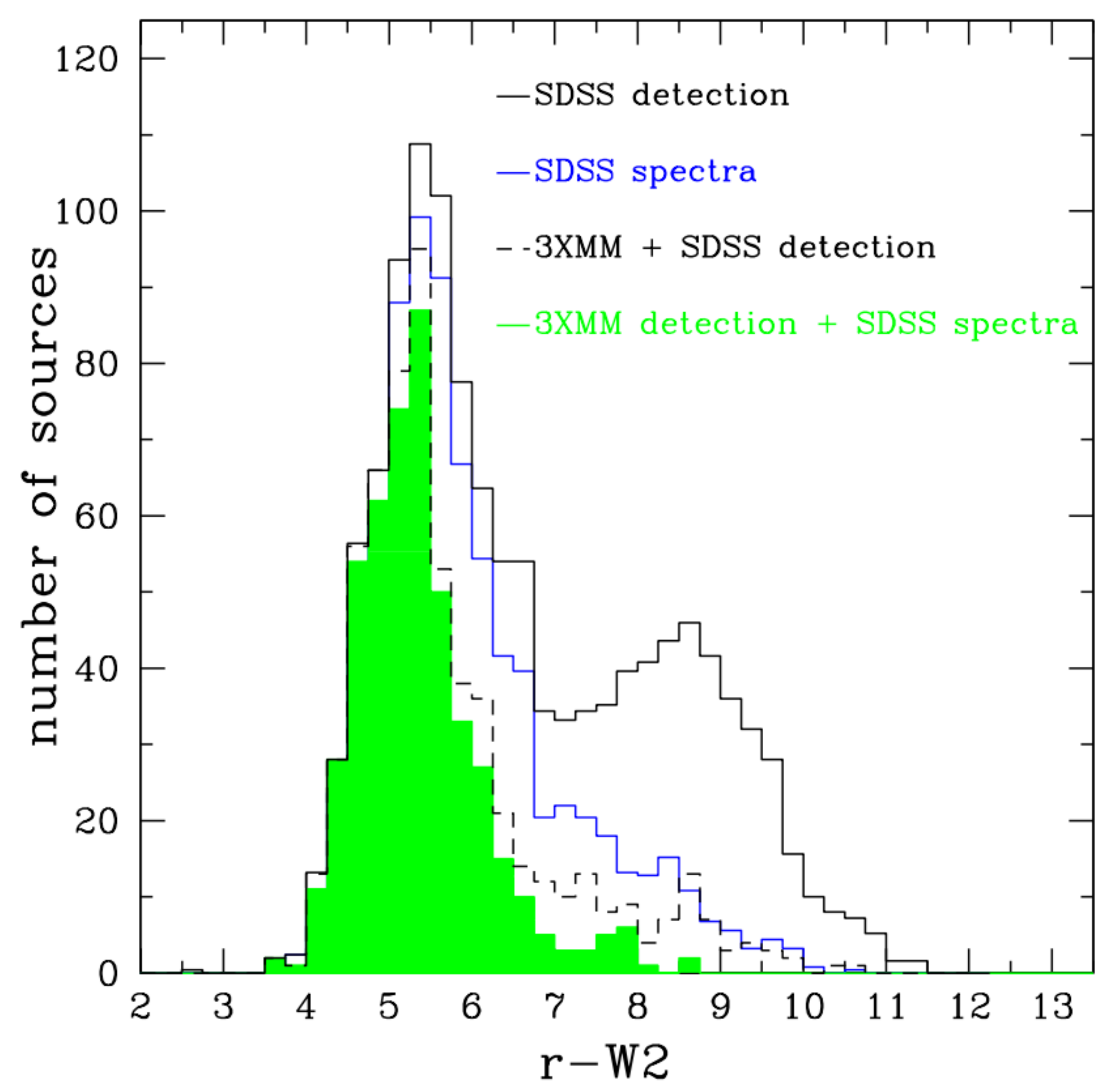}
  \label{fig_r_W2_distrib}
\end{subfigure}
\begin{subfigure}
  \centering
  \includegraphics[width=0.32\linewidth]{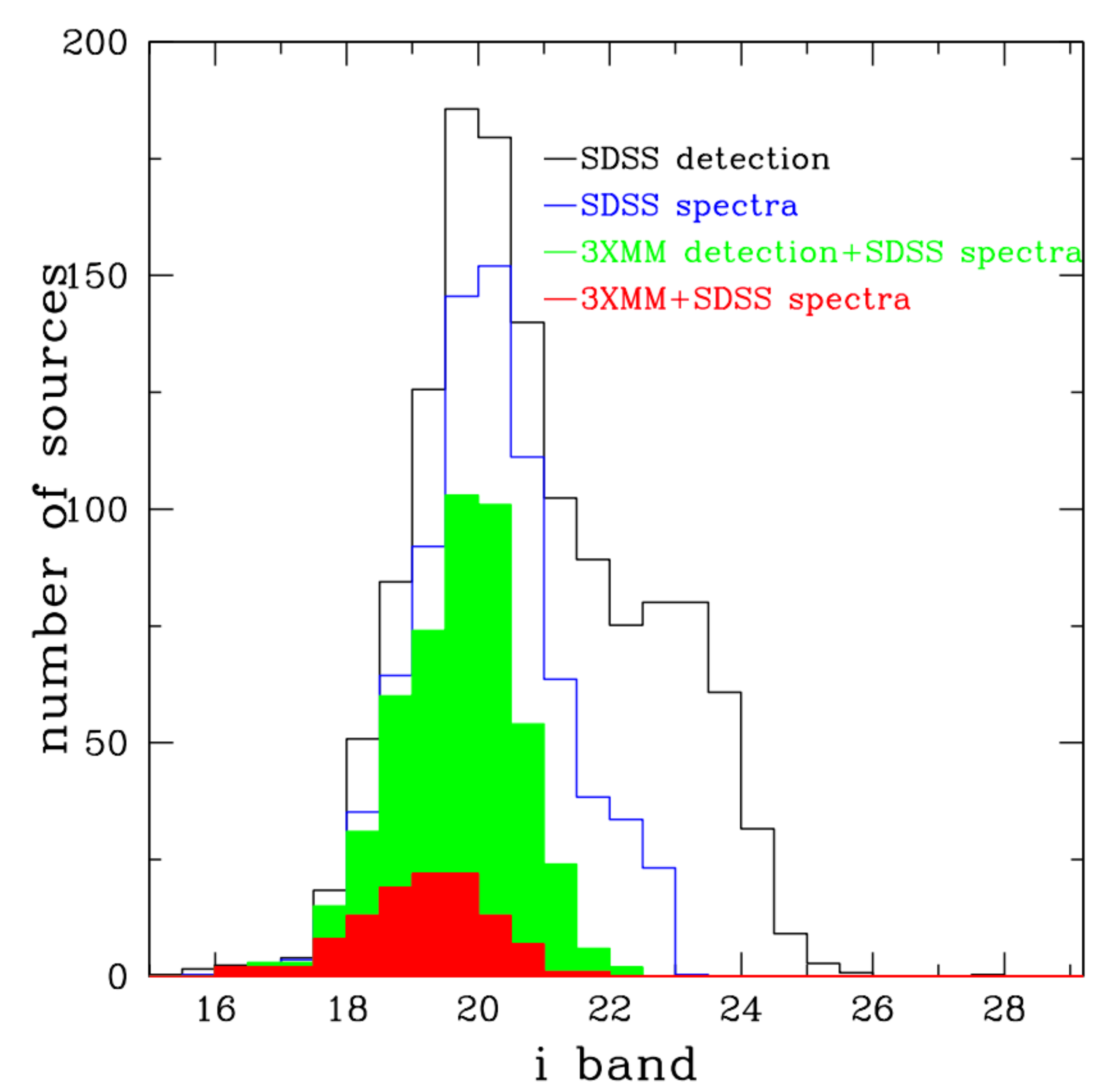}
  \label{fig_i_distrib}
\end{subfigure}
\begin{subfigure}
  \centering
  \includegraphics[width=0.32\linewidth]{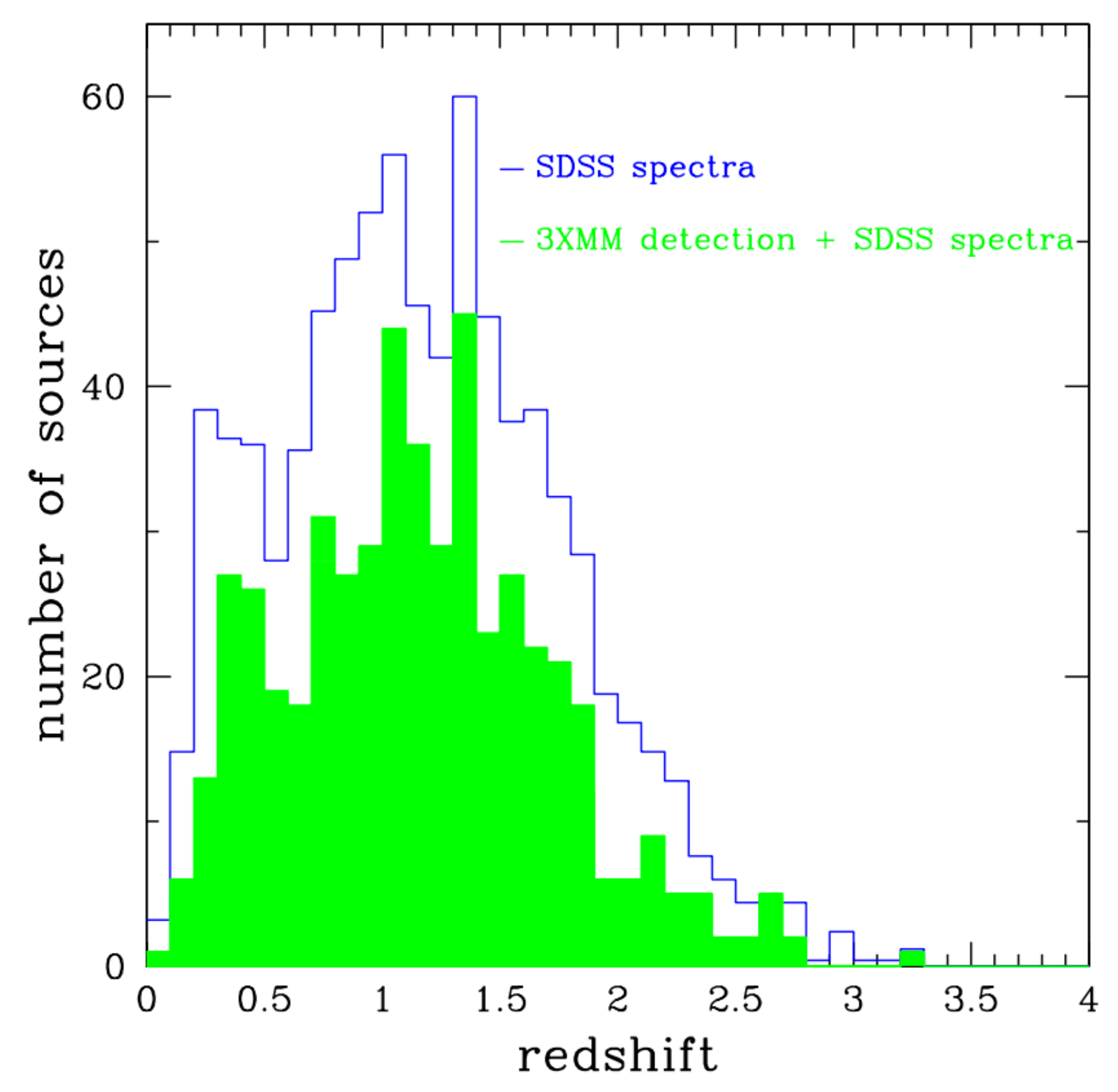}
  \label{fig_redz_distrib}
\end{subfigure}
\caption{From left to right: $\rm{r-W2}$, i-band and redshift distributions of the AGN samples.}
\label{redz_lx}
\end{figure*}

\section{Data}
\subsection{Sample selection}
\label{sec_selection}

Stripe 82 covers an area of $\sim$300\,deg$^2$ on the celestial equator and has been repeatedly scanned by the SDSS. In addition the field has been partially covered by observations at other wavelengths, e.g., UKIDSS \citep{Lawrence2007} and {\it{Herschel}} \citep{Viero2014}. The Stripe 82X survey \citep{LaMassa2013a, LaMassa2013b, LaMassa2015} covers $\sim$31\,deg$^2$ on the sky and was designed to take advantage of the wealth of multi-wavelength information provided by Stripe 82. Its goal is to reveal high luminosity AGN at high redshifts. 

Stripe 82X has both Chandra and XMM-{\it{Newton}} observations. In this work we study the X-ray spectral properties of IR selected AGN and compare them with their optical colours and SEDs. For that purpose we require 3XMM observations (Stripe 82-XMM). About half of the Stripe 82-XMM survey (15.6\,deg$^2$) is contiguous and reaches an $0.5-10$\,keV flux limit of $\rm \sim 10^{-14} \,erg\,s^{-1}\,cm^{-2}$ at half survey area. The median exposure time is $\sim 6$\,ks, while the coadded depth in theoverlapping regions reaches $\sim 6-8$\,ks \citep[][]{Lamassa2019}. The remaining 10.6\,deg$^2$, comprise of 4.6\,deg$^2$ of proprietary {\it{XMM-Newton}} data, with exposure time of $4-5$\,ks (12\,ks in regions with greatest overlap) and 6\,deg$^2$ of archival data \citep[][]{Lamassa2013}.



We use $\sim300, 000$ {\it{WISE}} sources in the Stripe 82-XMM area to select mid-IR AGN. We apply the criteria presented in \cite{Stern2012} combined with the modified selection criterion described in \cite{Assef2013}, for fainter {\it{WISE}} sources. Specifically, the selection criteria we apply are:\\
For $\rm{W2}<15.05$
\begin{equation}
\rm{W1-W2} \geq 0.8,
\end{equation}
and for $15.05\leq \rm{W2}\leq 17.11$
\begin{equation}
\rm{W1-W2}>a_R\,\rm{exp}\,[\beta _R (\rm{W2}-\gamma _R)^2],
\\
W1_{snr}>3,
\end{equation}
where W1, W2 are the {\it{WISE}} photometric bands, at 3.4 and 4.6$\mu$m,  W1$_{snr}$ is the S/N ratio of the W1 band and $(a_R, \beta _R, \gamma _R)=(0.662, 0.232, 13.93)$, to select AGN at a 90\% reliability.  We find 1946 IR AGN that lie within the 3XMM footprint. 1507 of these sources have optical \citep[SDSS DR13;][]{Albareti2017} counterpart and 824 have available optical spectrum. We cross-match these sources with the 499, 266 X-ray sources of the 3XMM-DR7 catalogue \citep{Rosen2016}. For the cross-match we use TOPCAT, version 4.6, using a radius of 3 arcsec. 0.2\% spurious sources are expected among common sources between the two datasets \citep{Mountrichas2017a}. The number of sources in each subsample is presented in Table \ref{table_data}. 1250 (1946-696), i.e., $\sim{64\%}$, of the IR selected AGN are undetected in X-rays. This percentage is in agreement with the $58\%$ found in \cite{Lamassa2019}.

\subsection{Sample properties}

The scope of the paper is to study the X-ray absorption of IR selected AGN and compare it with optical/mid-IR criteria. Therefore, both X-ray and optical identification are needed for the sources. However, this requirement imposes selection biases in the AGN sample. In the following, we study the impact of these biases on our X-ray sample.

Previous studies have found a correlation between the optical colour and the X-ray obscuration \citep[e.g.,][]{Civano2012}. \cite{Yan2013} showed that type-2 AGN candidates at z$\leq 3$, can be identified by applying the following colour criteria: {\it{WISE}} $\rm{W1-W2}>0.8, \rm{W2}<15.2$ combined with $\rm{r-W2}>6$ (Vega). In Fig. \ref{fig_r_W2_distrib} (left panel) we plot the $\rm{r-W2}$ distributions of the AGN samples. Sources with only SDSS detection, show a peak in their distribution at $\rm{r-W2}\approx 5.5$. However, a second peak is present at larger $\rm{r-W2}$ values ($\approx 8.5$). When an SDSS spectrum requirement is applied, this second peak is not detected. However, we notice a wide tail at high r-W2 values. Our X-ray detected sources completely miss this optically, possibly obscured AGN population. These distributions are quantified in Table \ref{table_yan}, that presents the fraction of optically red sources in the various subsamples. Based on the $\rm{r-W2}$ distributions, we notice that $\rm{r-W2}=7$ cut seems more natural, since it separates two peaks. Given the prevalence of the  $\rm{r-W2}=6$ cut in the literature, we keep the latter cut in our analysis. However, in Table \ref{table_yan} we also present the fraction of optically red sources, using the $\rm{r-W2}=7$ cut. Regardless of the exact value of the optical/mid-IR cut applied, the measurements confirm our aforementioned findings, i.e., IR selected AGN that are undetected by X-rays are probably preferentially redder than those detected in X-rays \citep[but see Fig. 16 in][]{Lamassa2019}. Middle and right panels of Fig. \ref{fig_i_distrib}, present the i-band and redshift distributions of the AGN samples. We notice that X-ray detected sources are brighter than i$\lessapprox 22$.

\cite{Mateos2012} used MIR {\it{WISE}} colours to select luminous AGN candidates. Using these AGN, they defined a highly complete, mid-IR colour wedge that their sources occupy. X-ray detected AGN reside in the lower part of this wedge (see their Figures 2 and 6). In Fig. \ref{fig_w1w2w3_distrib} we plot W1-W2 versus W2-W3 colours, for those of our sources with $W3_{snr}>5$. In agreement with  \cite{Mateos2012}, our X-ray detected AGN occupy the lower part of the mid-IR colour diagram. However, optically red sources ($r-W2>6$, red points in Fig. \ref{fig_w1w2w3_distrib} ) seem to be randomly distributed in the colour diagram.


\begin{table}
\caption{The number of  mid-IR selected AGN in Stripe 82X, with optical/X-ray counterparts and spectra.}
\centering
\setlength{\tabcolsep}{0.3mm}
\begin{tabular}{cccccc}
      \hline
 Mid-IR selected AGN & 1946 \\ 
 3XMM detection  & 696 \\
SDSS detection & 1507  \\
SDSS spectra & 824 \\
3XMM and SDSS detection & 646 \\ 
3XMM detection + SDSS spectra &  507 \\
       \hline
\hline
\label{table_data}
\end{tabular}
\end{table}

\begin{table}
\caption{Fraction of optically red sources found in our IR selected AGN in Stripe 82X. In the parentheses we quote the total number of AGN in each subsample.}
\centering
\setlength{\tabcolsep}{2.mm}
\begin{tabular}{lcc}
      \hline
sample & red sources& red sources\\
 & $\rm r-W2>6$ & $\rm r-W2>7$ \\
 \hline
SDSS detection (1507) & 43\% & 30\%\\
SDSS spectra (824) & 31\% & 13\%\\
3XMM+SDSS detection (646) & 23\% & 9\% \\
3XMM detection + SDSS spectra (507) & 20\% &  6\%\\   
\hline
\hline
\label{table_yan}
\end{tabular}
\end{table}

\begin{figure*}
\centering
\begin{subfigure}
  \centering
  \includegraphics[width=0.43\linewidth]{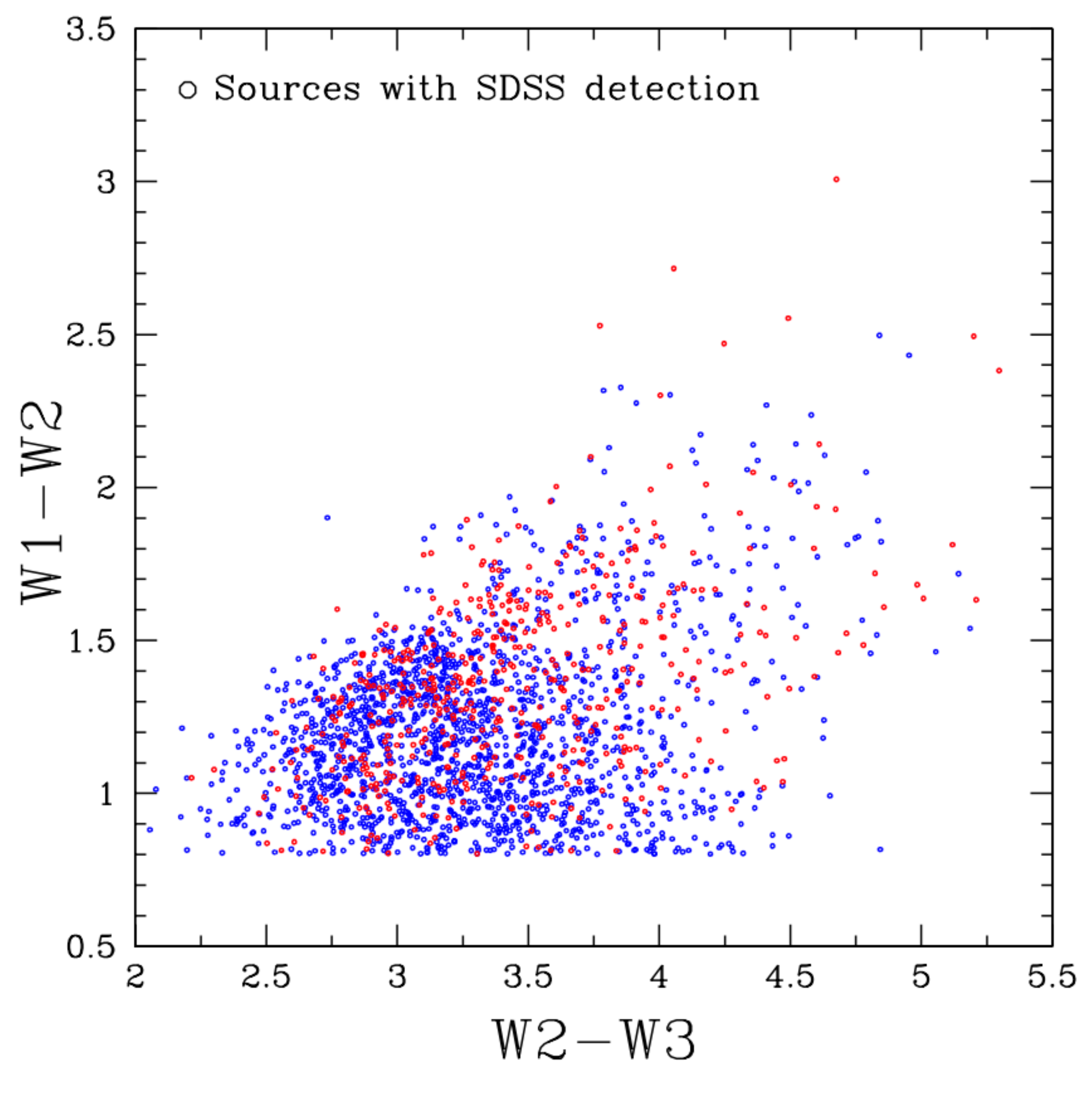}
  \label{fig_w1w2w3_distrib}
\end{subfigure}
\begin{subfigure}
  \centering
  \includegraphics[width=0.445\linewidth]{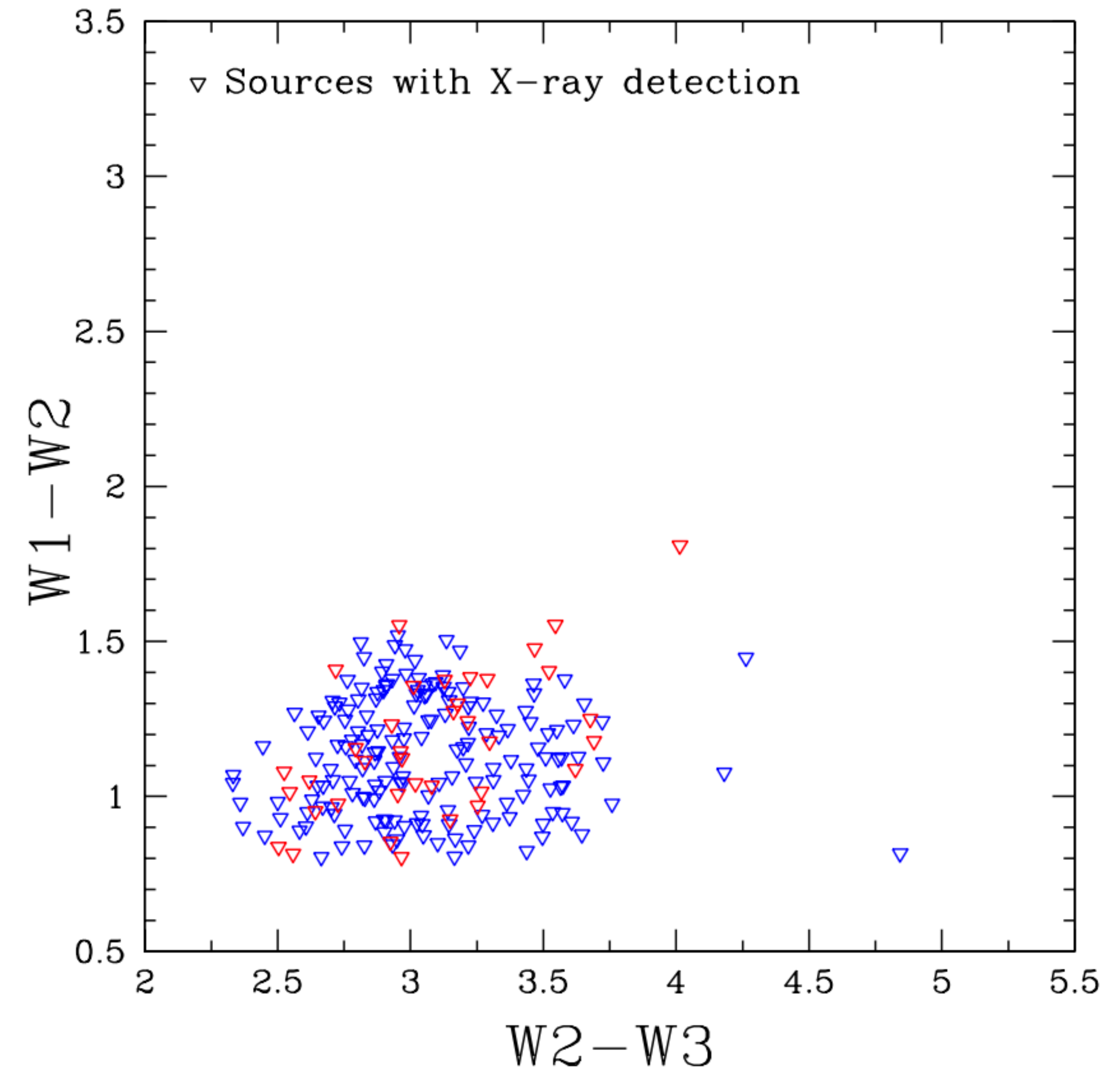}
  \label{fig_w1w2w3_distrib}
\end{subfigure}
\caption{W1-W2 vs. W2-W3 of those sources with SDSS detection (left panel) and those with X-ray detection (right panel). Sources with W3$_{\rm snr}<5$ have been excluded. Optically blue sources (r-W2$<$6) are shown in blue. Optically red sources are shown in red (r-W2$>$6).}
\end{figure*}

\begin{figure*}
\centering
\begin{subfigure}
  \centering
  \includegraphics[width=0.484
  \linewidth]{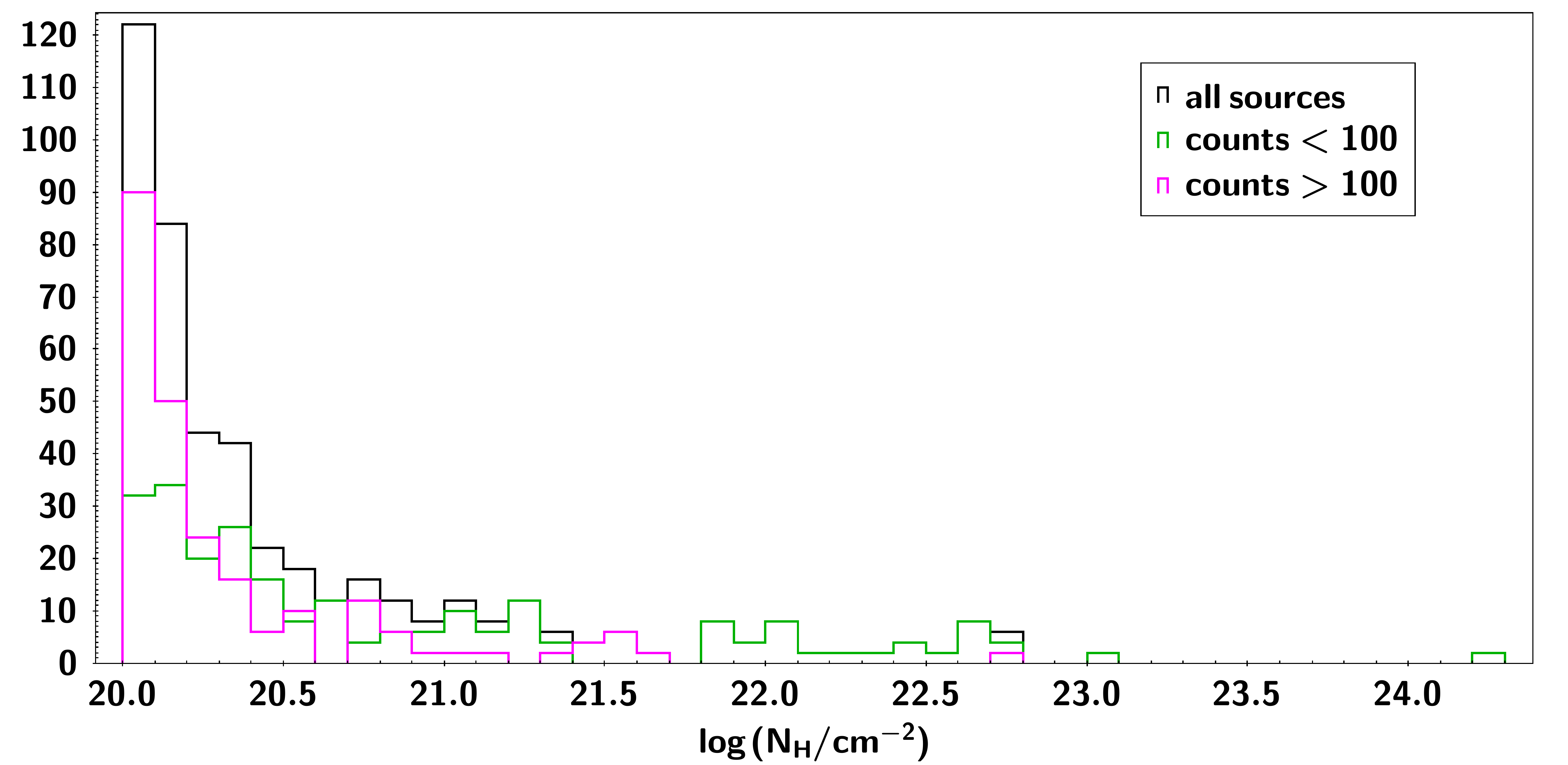}
\end{subfigure}
\begin{subfigure}
  \centering
  \includegraphics[width=0.44\linewidth]{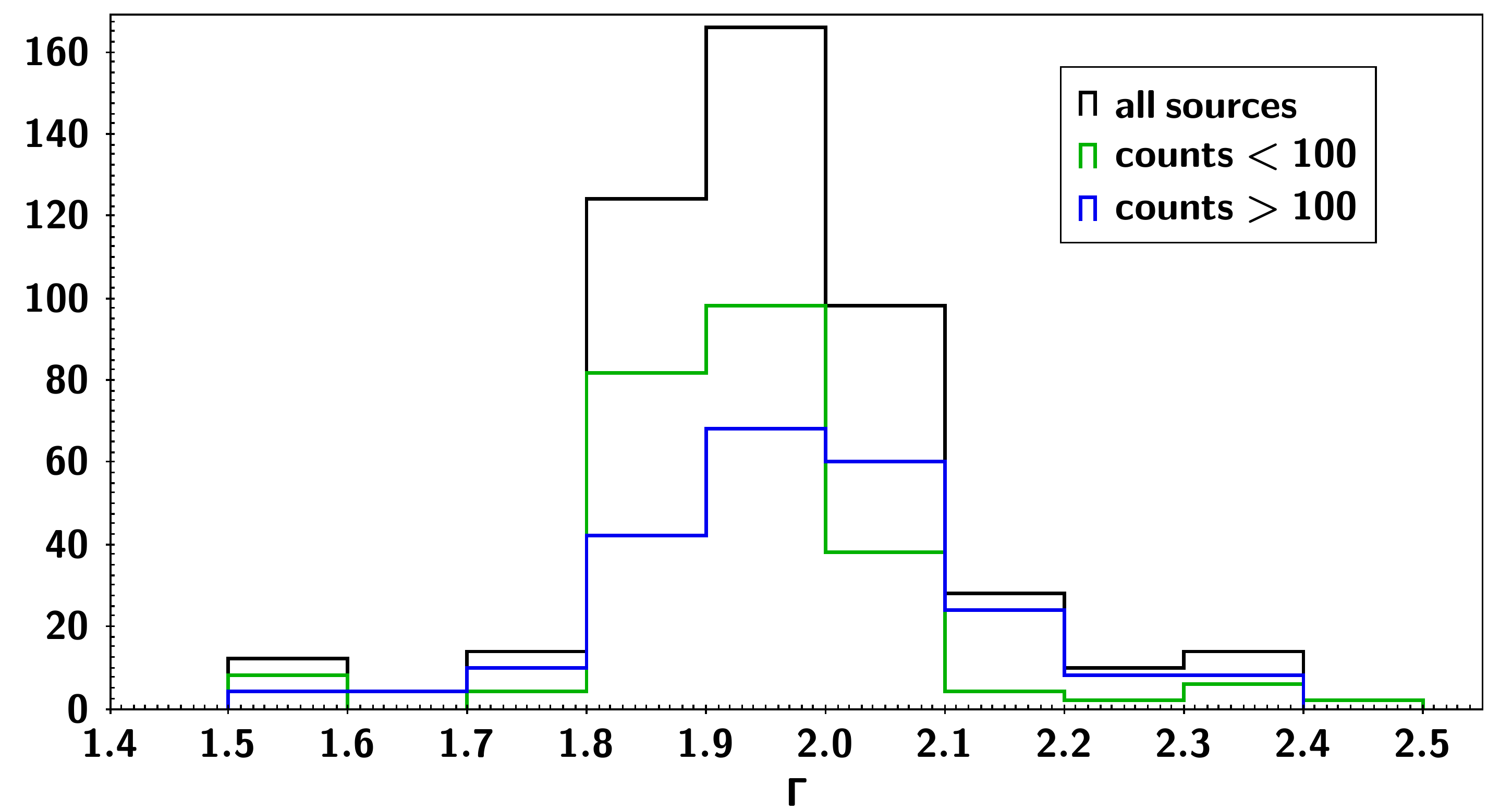}
\end{subfigure}
\caption{The distribution of N$_H$ (left panel) and $\Gamma$ (right panel) values of the 507 IR AGN with X-ray detection (black line), using spectral fitting. The green line presents the distributions of those sources with less than 100 photon counts and the blue line the distributions of those sources with more than 100 counts.}
\label{fig_nh_distrib}
\end{figure*}

\begin{table}
\caption{The mean values of $\rm N_H$ and $\Gamma$ for the X-ray sample as well as for various subsamples, based on the number of source photons. The errors correspond to the mean values of the $1\sigma$ uncertainties of the measurements.}
\centering
\setlength{\tabcolsep}{2.mm}
\begin{tabular}{lccc}
      \hline
sample & no. of sources &$\rm log\,( N_H / cm^{-2}) $&$\Gamma$\\
\hline
All sources & 507 & $20.6\pm 0.6$ & $1.96\pm 0.20$ \\
$<50$ counts & 152 & $20.9\pm 0.9$ & $1.93\pm 0.28$ \\
$<100$ counts & 250 & $20.9\pm 0.8$ & $1.93\pm 0.25$ \\
$>100$ counts & 257 & $20.3\pm 0.4$ & $1.99\pm 0.17$ \\
$>500$ counts & 40 & $20.2\pm 0.3$ & $2.05\pm 0.09$ \\
\hline
\hline
\label{table_photons}
\end{tabular}
\end{table}


\section{X-ray and optical absorption of IR AGN}
\label{sec_xray}

In this Section, we study the X-ray absorption of the 507 IR selected AGN with X-ray detection and optical spectra. We then compare the X-ray with the optical absorption \citep[e.g.,][]{Glikman2018}, using optical/mid-IR colours and SED fitting.

To study the X-ray properties of the IR selected AGN,  we use Xspec v12.8 \citep{Arnaud1996}. Photoelectric absorption is included in all the spectral models (wabs, in XSPEC notation), fixed at the Galactic value at the source coordinates given by Leiden/Argentine/Bonn (LAB) Survey of the Galactic HI \citep{Kalberla2005}. A single power-law model absorbed by neutral material surrounding the central source is applied on the X-ray spectra. For that purpose, we used the fitting and modelling software Sherpa \citep[version 4.9.1;][]{Freeman2001}. We followed the Bayesian technique proposed in \cite{Buchner2014}, using the analysis software BXA (Bayesian X-ray Analysis), that connects the nested sampling algorithm MultiNest \citep[][]{Feroz2009} with Sherpa. A nested sampling algorithm allows a full exploitation of the parameter space, avoiding solutions that correspond to local minima which is a common problem with standard minimization techniques (e.g. the Levenberg-Marquardt algorithm). In the BXA framework, a probability prior is assigned to each free parameter of the model. For the power law, photon index, $\Gamma$, we used a gaussian prior with mean value 1.9 and standard deviation 0.15 \citep{Nandra1994}. For the remaining parameters, we used flat, uninformative priors. Our spectral fits, follow the same procedure to build the XMMFITCAT-z (see Ruiz et al., in prep., http://xraygroup.astro.noa.gr/Webpage-prodex/).

The mean N$_H$ and $\Gamma$ values of the sample are $\rm N_H=10^{20.6\pm0.6}~cm^{-2}$ and $\Gamma = 1.96\pm0.20$. The quoted errors correspond to the mean values of the 1\,$\sigma$ uncertainties. We also split the sample into various subsamples, based on the number of source photons. The mean values of $\rm N_H$ and $\Gamma$ are presented in Table \ref{table_photons}. Fig. \ref{fig_nh_distrib} presents the distributions of N$_H$ (left panel) and $\Gamma$ (right panel) values. Our analysis reveals 40 sources ($\sim 8\%$) that have $\rm N_H>10^{22}~cm^{-2}$. This number increases to 65 ($\sim 13\%$) if we loosen our X-ray absorption criterion, i.e., $\rm N_H>10^{21.5}~cm^{-2}$. However, we note that our X-ray spectra extend to energies up to 10\,keV. Modelling X-ray spectra at higher energies could possibly change the estimated $\rm N_H$ values \citep{Civano2015}.

\begin{figure}
\centering
  \centering
  \includegraphics[height=1.\columnwidth]{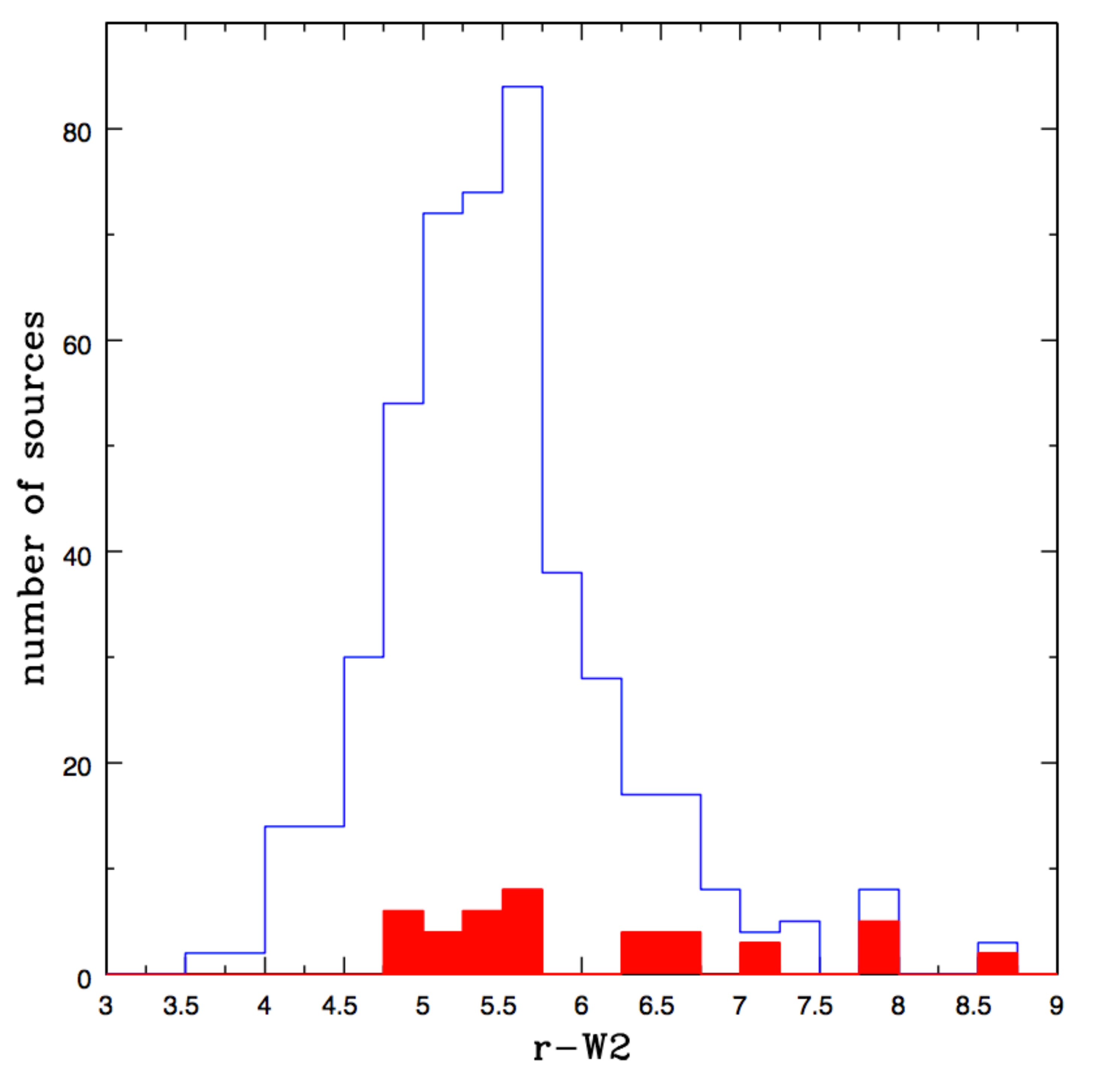}
\caption{$\rm{r-W2}$ distributions of the 507 IR AGN with X-ray detection (blue line) and the 40/507 that are X-ray absorbed (red shaded region).}
\label{fig_r_w2_nh_distrib}
\end{figure}

\begin{figure*}
\centering
\begin{subfigure}
  \centering
  \includegraphics[width=0.45\linewidth]{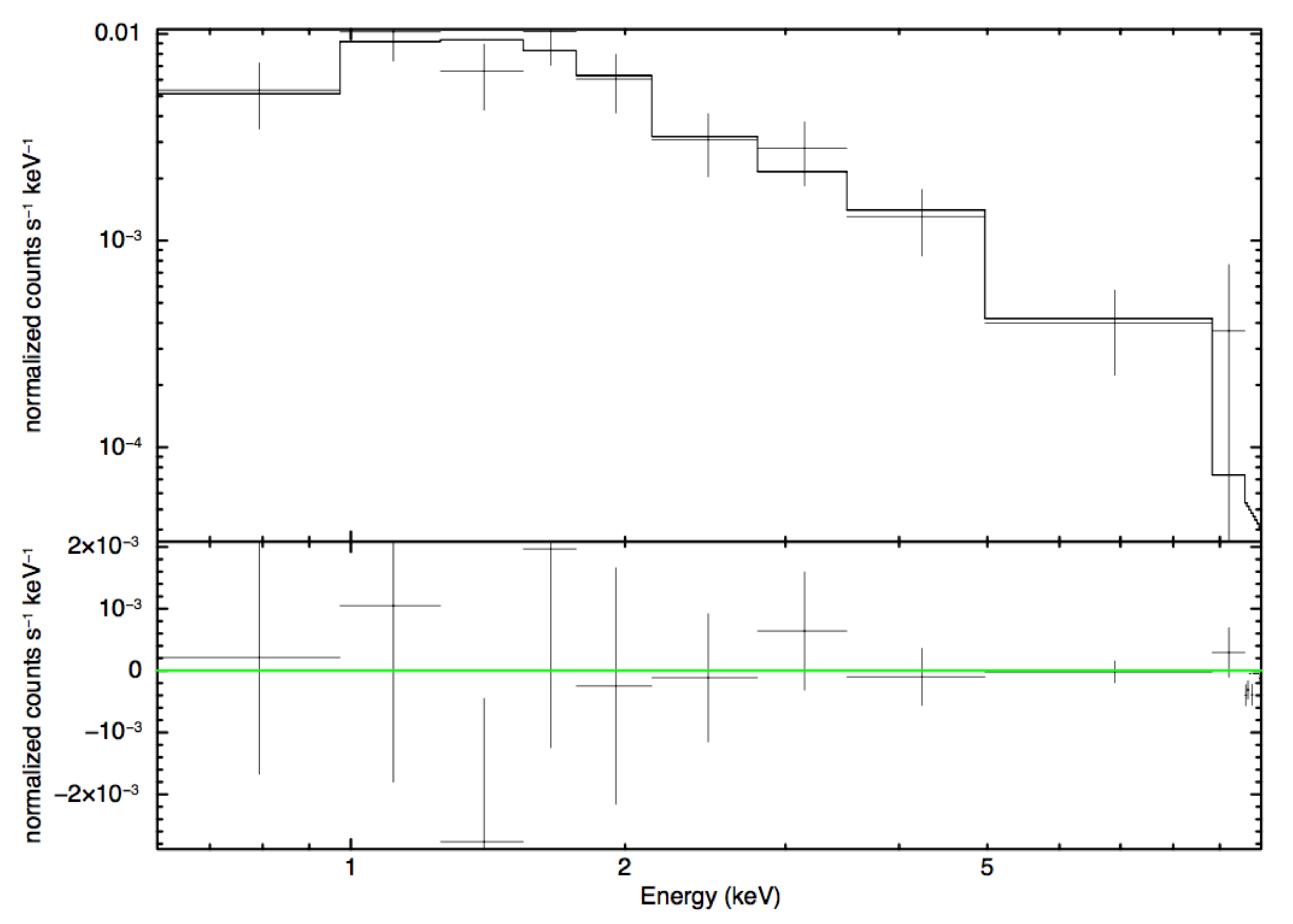}
  \label{}
\end{subfigure}
\begin{subfigure}
  \centering
  \includegraphics[width=0.44\linewidth]{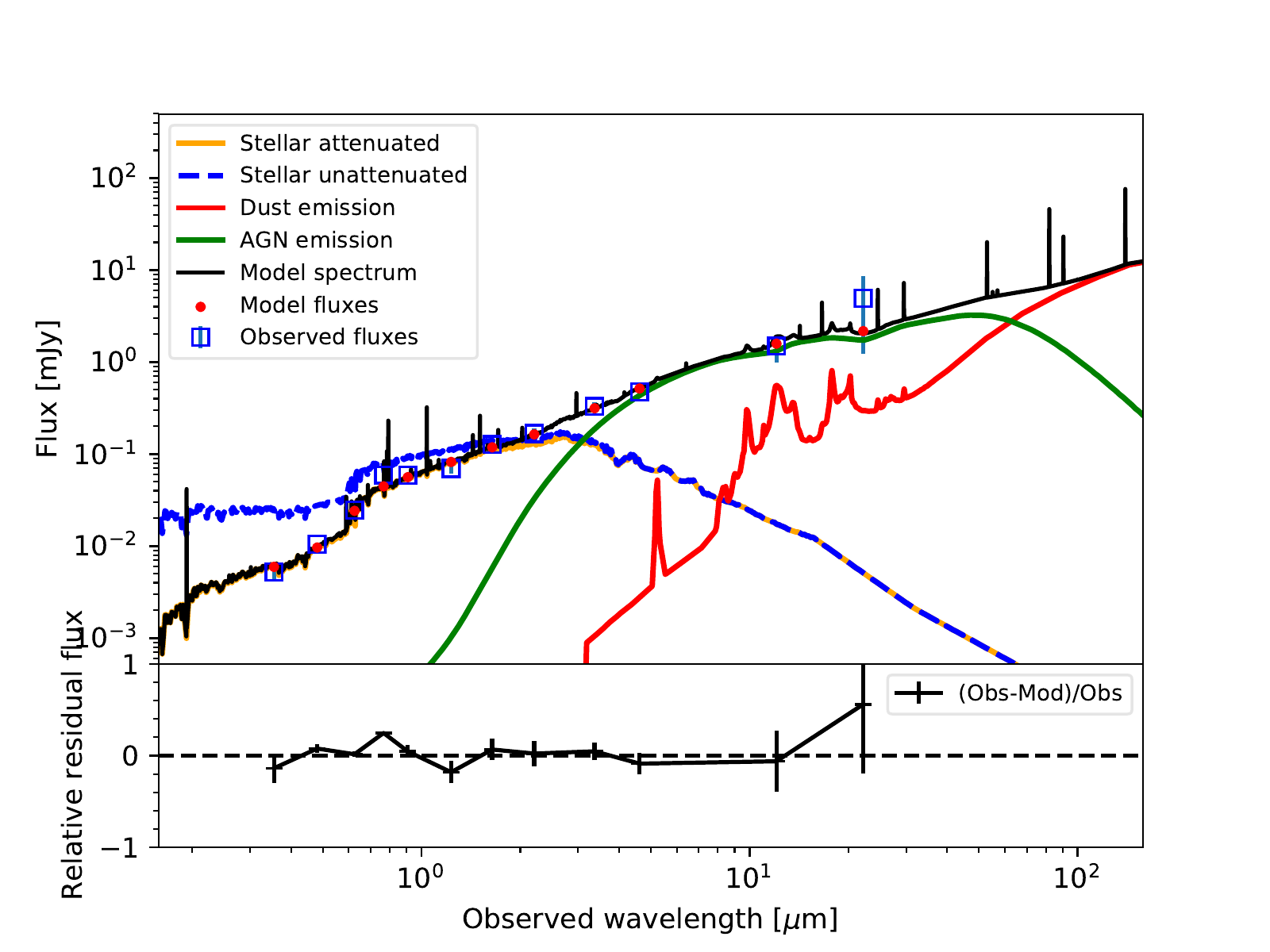}
  \label{}
\end{subfigure}
\caption{{\bf{An example of an IR AGN at $\rm ra=17.55, dec=-0.13$, $\rm z=0.58$, that is X-ray absorbed (log\,(N$_H/\rm{cm}^{-2})=22.34^{+0.13}_{-0.22}$, $\Gamma = 2.12 ^{+0.25}_{-0.09}$)}}, red ($r-W2=7.46$) and optically obscured ($\Psi=0.001^{o}$).}
\label{x_abs_red_sed_abs}
\end{figure*}

\begin{figure*}
\centering
\begin{subfigure}
  \centering
  \includegraphics[width=0.45\linewidth]{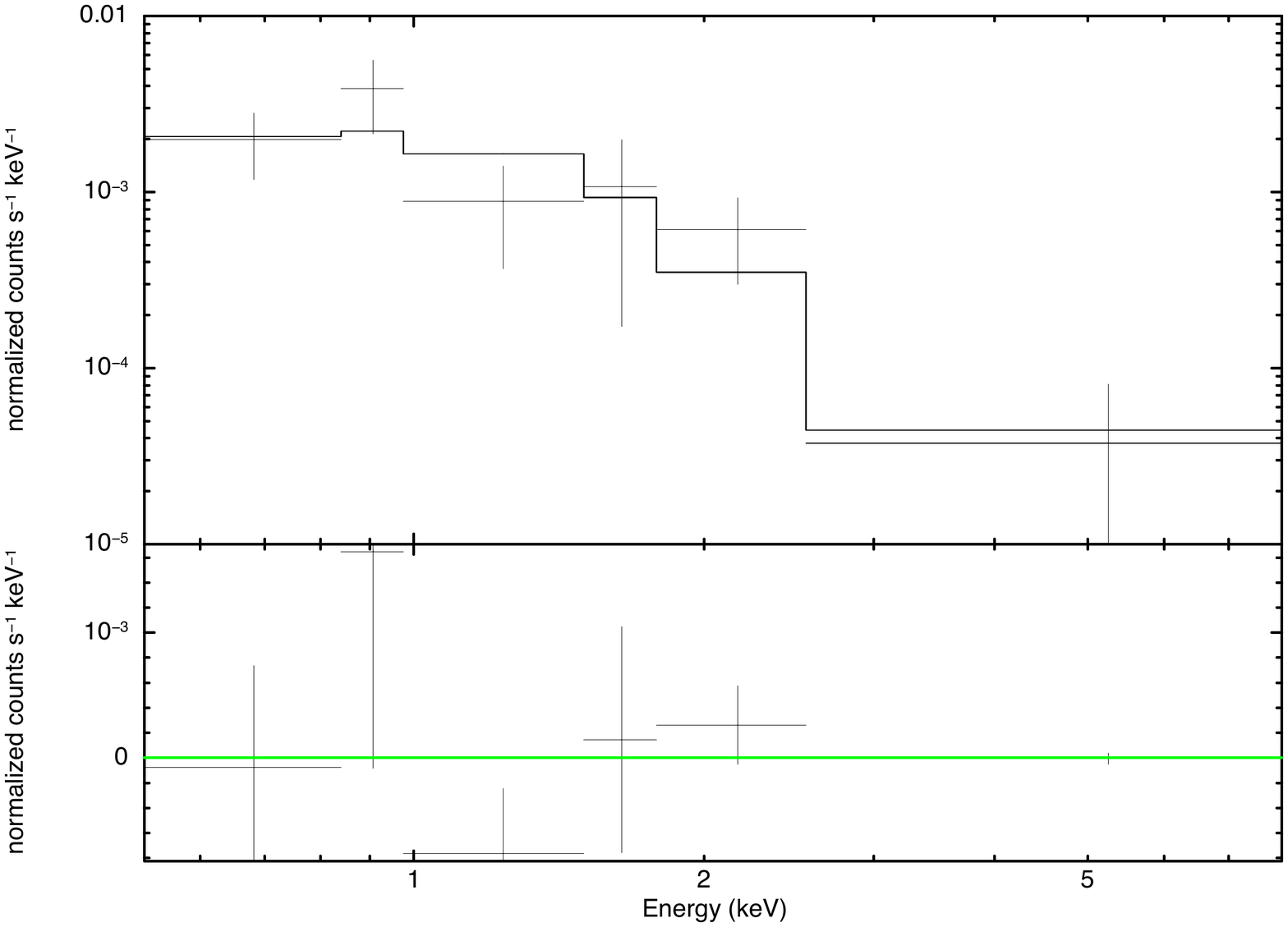}
  \label{}
\end{subfigure}
\begin{subfigure}
  \centering
  \includegraphics[width=0.44\linewidth]{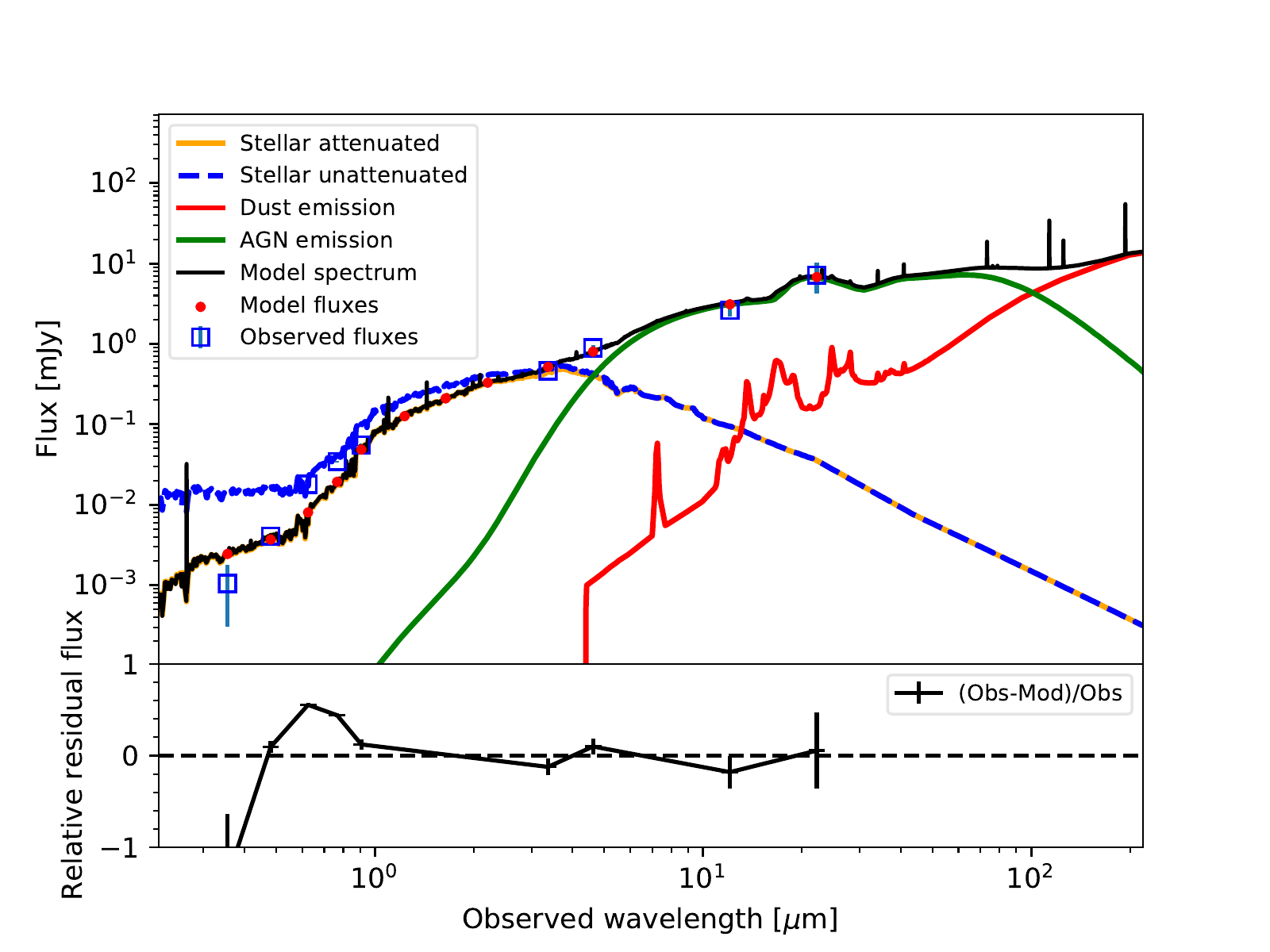}
  \label{}
\end{subfigure}
\caption{{\bf{An example of an IR AGN at $\rm ra=22.75, dec=0.29$, $\rm z=1.22$, that is X-ray unabsorbed (log\,(N$_H/\rm{cm}^{-2})=21.35^{+0.15}_{-0.06}$, $\Gamma = 1.62 ^{+0.13}_{-0.21}$)}}, red ($r-W2=9.16$) and optically obscured ($\Psi=0.001^{o}$).}
\label{x_unabs_red_sed_abs}
\end{figure*}

\begin{figure*}
\centering
\begin{subfigure}
  \centering
  \includegraphics[width=0.45\linewidth]{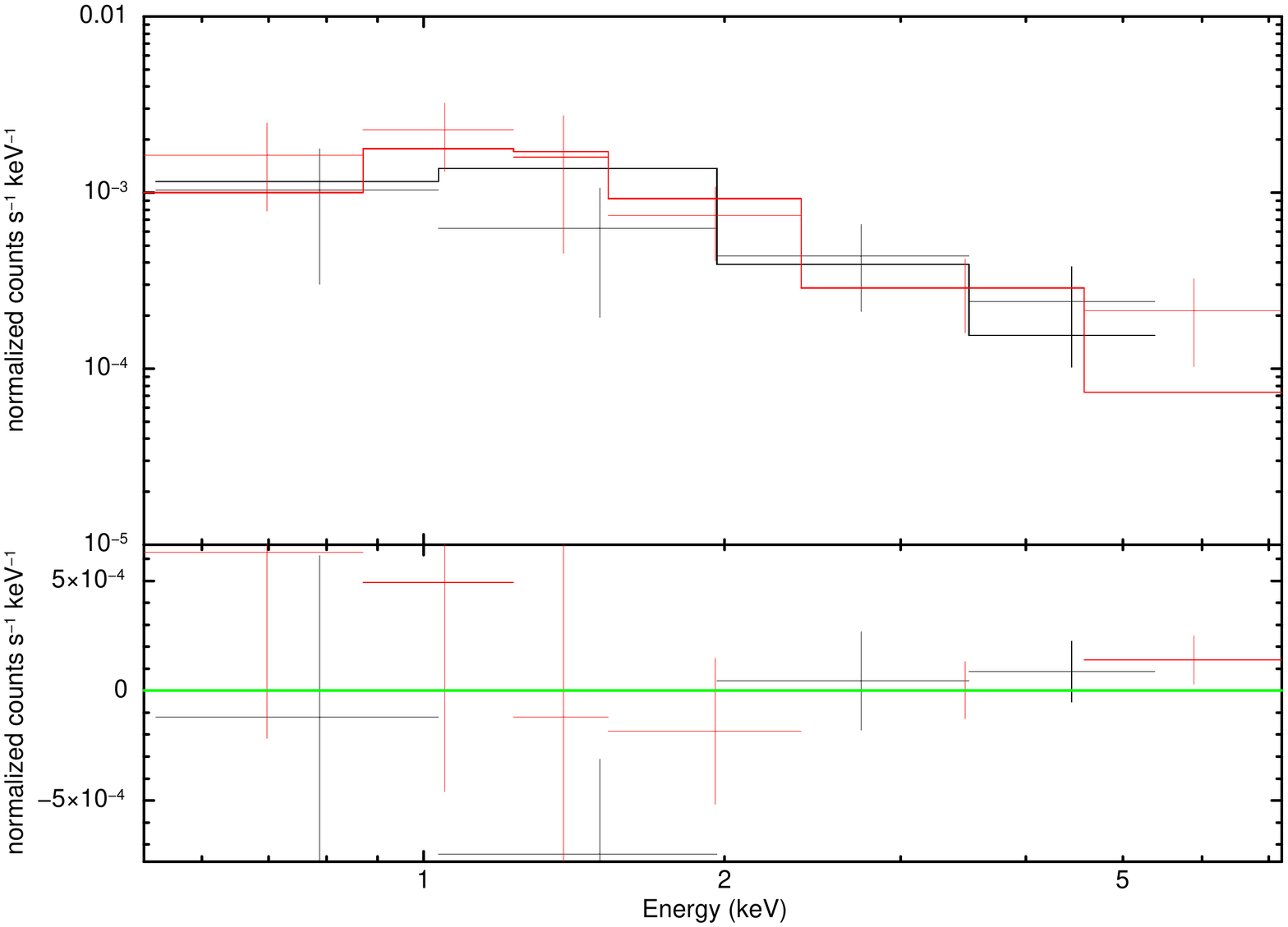}
  \label{}
\end{subfigure}
\begin{subfigure}
  \centering
  \includegraphics[width=0.44\linewidth]{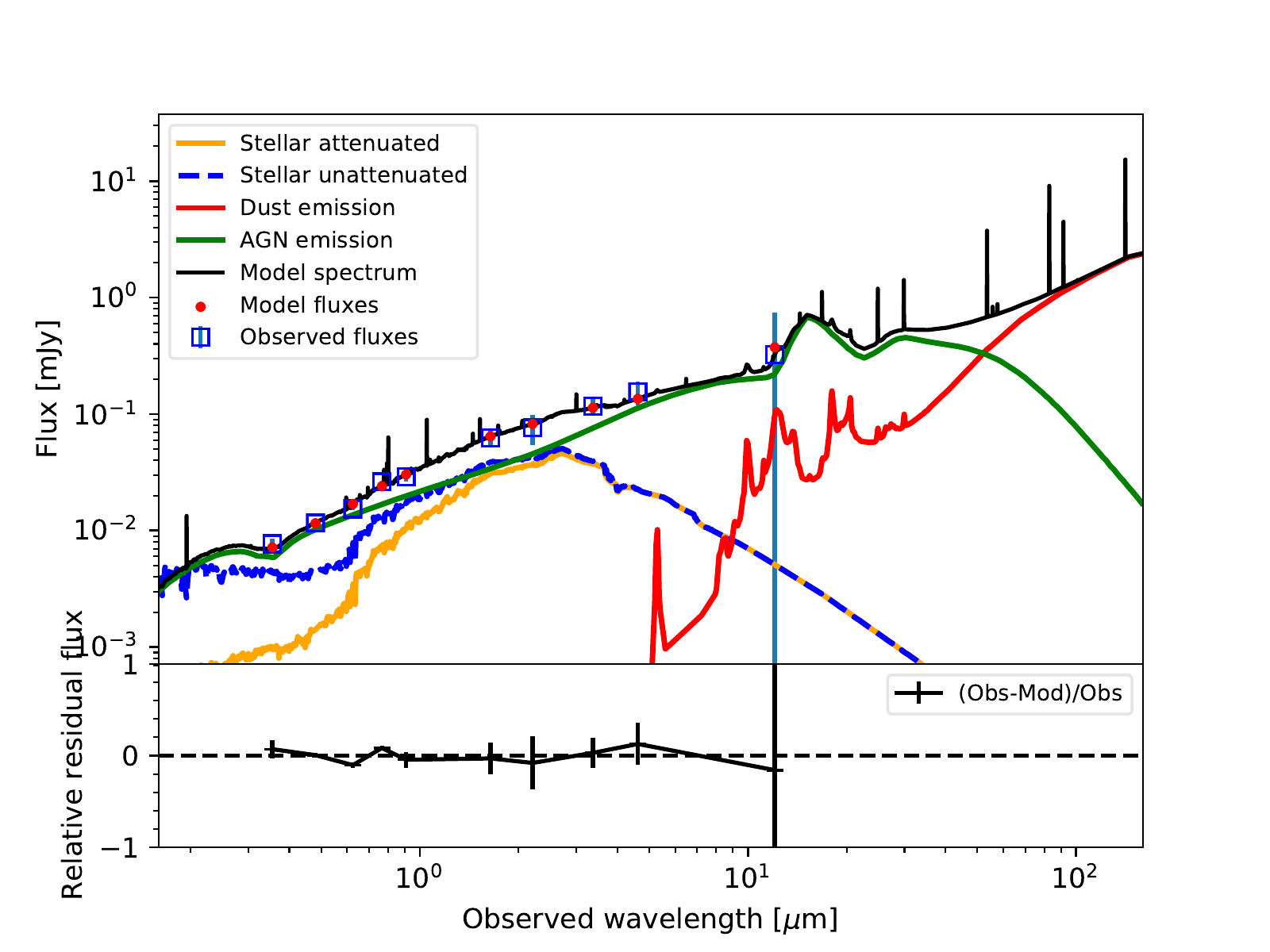}
  \label{}
\end{subfigure}
\caption{{\bf{An example of an IR AGN at $\rm ra=26.57, dec=0.87$, $\rm z=1.3$, that is X-ray unabsorbed (log\,(N$_H/\rm{cm}^{-2})=21.67^{+0.25}_{-0.19}$, $\Gamma = 1.75 ^{+0.17}_{-0.12}$)}}, red ($r-W2=6.23$) and optically unobscured ($\Psi=89.99^{o}$). The SED reveals that the source is a galaxy dominated system (see text for more details).}
\label{x_unabs_red_sed_unabs}
\end{figure*}

Optical/mid-IR colour criteria reveal that $18/40$ (45\%) of the X-ray absorbed AGN are also optically red ($\rm{r-W2}>6$). This percentage is higher compared to the fraction of red sources among the 507 sources (20\%, see Table \ref{table_yan}). However, among the non X-ray absorbed sources (467), there are 70 (15\%) optically red AGN (Fig. \ref{fig_r_w2_nh_distrib}). Prompted by this, we construct SEDs for the sample of the 507 AGN and further examine their obscuration (for more details see Appendix). The SED fitting analysis reveals that 86\% of the red objects are optically obscured, based on the estimated inclination angle of the torus. Thus, there is a very good agreement between the optical colours and the SEDs of the sources. All 18 IR AGN that are X-ray absorbed and optically red are also obscured based on their SEDs. An example of the X-ray spectrum and the SED of one of these sources is presented in Fig. \ref{x_abs_red_sed_abs}. Among the 70 optically red AGN that are not X-ray absorbed, 58 (83\%) are obscured, based on their SEDs (Fig. \ref{x_unabs_red_sed_abs}). The $\rm r-W2$ values and the X-ray spectral results of these sources are presented in Table \ref{table_58_sources}. As can be seen in this Table, there are a few sources that present some X-ray obscuration, but do not meet our X-ray obscuration criterion, i.e., their $\rm N_H$ value is $(10^{21.5} < \rm N_H <10^{22})~cm^{-2}$. However, a large number of sources that have high number of counts (a few hundred up to a couple of thousand) present no indication of X-ray obscuration ($\rm N_H<10^{20.5-21}~cm^{-2}$). For, the remaining 12 sources (17\%) the SEDs reveal the they are galaxy dominated systems, i.e., the galaxy emission is a primary component of the SED in the optical part of the spectrum, even though the AGN emission is not absorbed at these wavelengths (Fig. \ref{x_unabs_red_sed_unabs}). The aforementioned numbers do not change if we lower our X-ray absorption criterion, i.e. $\rm N_H>10^{21.5}~cm^{-2}$. We conclude, the SED fitting analysis confirms that there is a fraction ($\frac{58}{467}\approx12\%$) of IR AGN that are red, obscured sources without presenting X-ray absorption. 

\begin{figure*}
\centering
\begin{subfigure}
  \centering
  \includegraphics[width=0.45\linewidth]{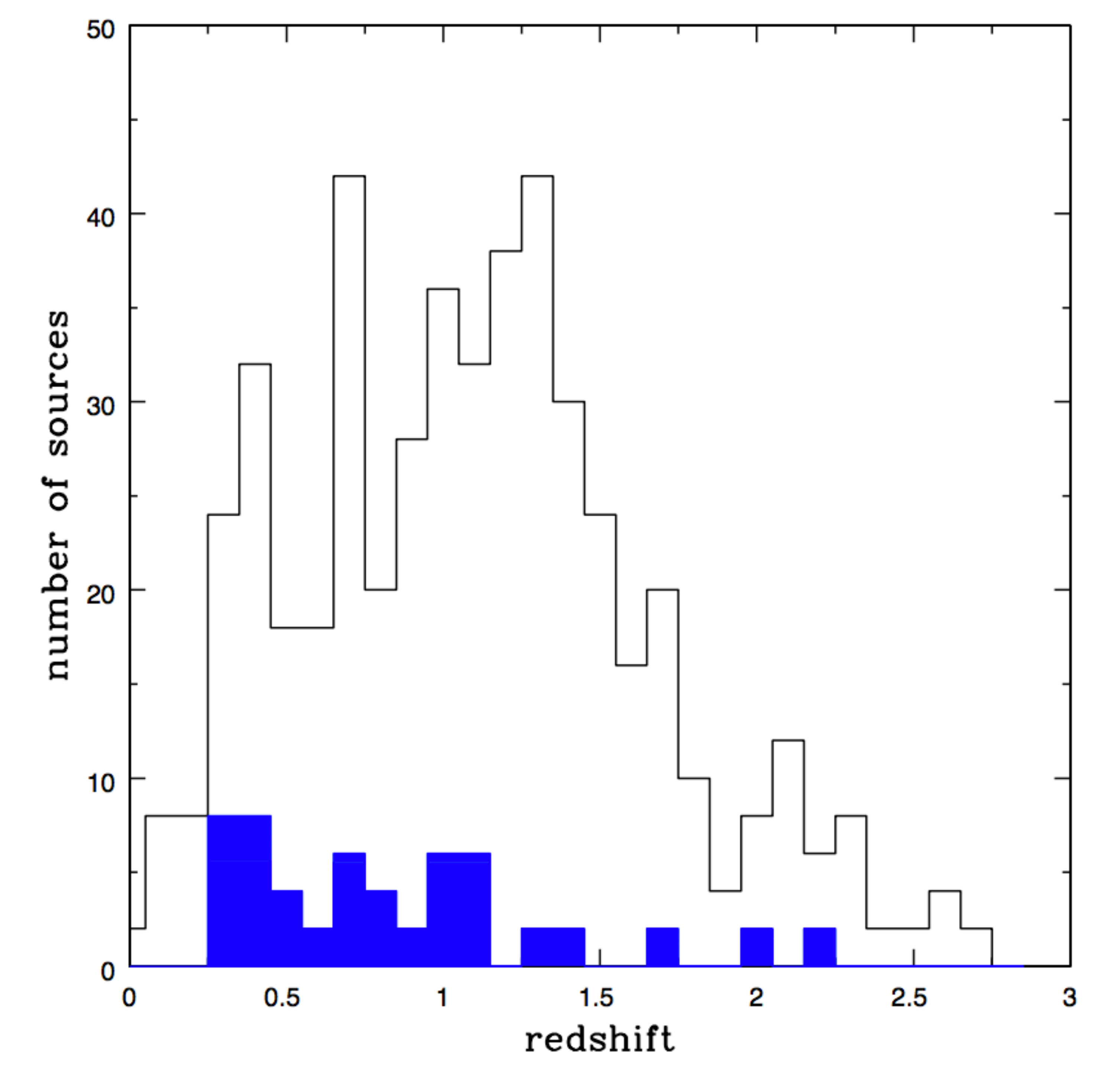}
  \label{fig_z_distrib_peculiar}
\end{subfigure}
\begin{subfigure}
  \centering
  \includegraphics[width=0.45\linewidth]{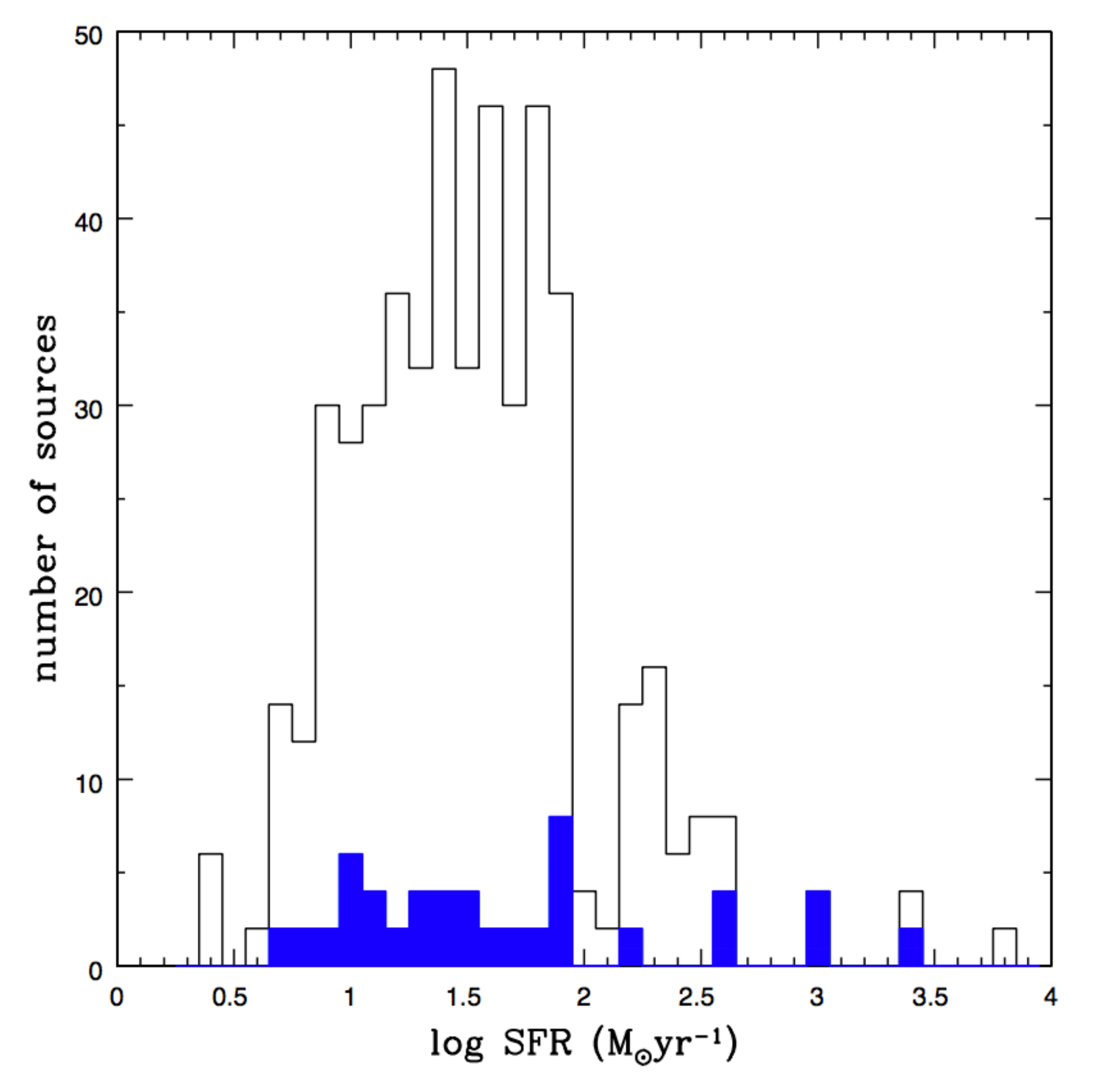}
  \label{fig_sfr_distrib_peculiar}
\end{subfigure}
\begin{subfigure}
  \centering
  \includegraphics[width=0.45\linewidth]{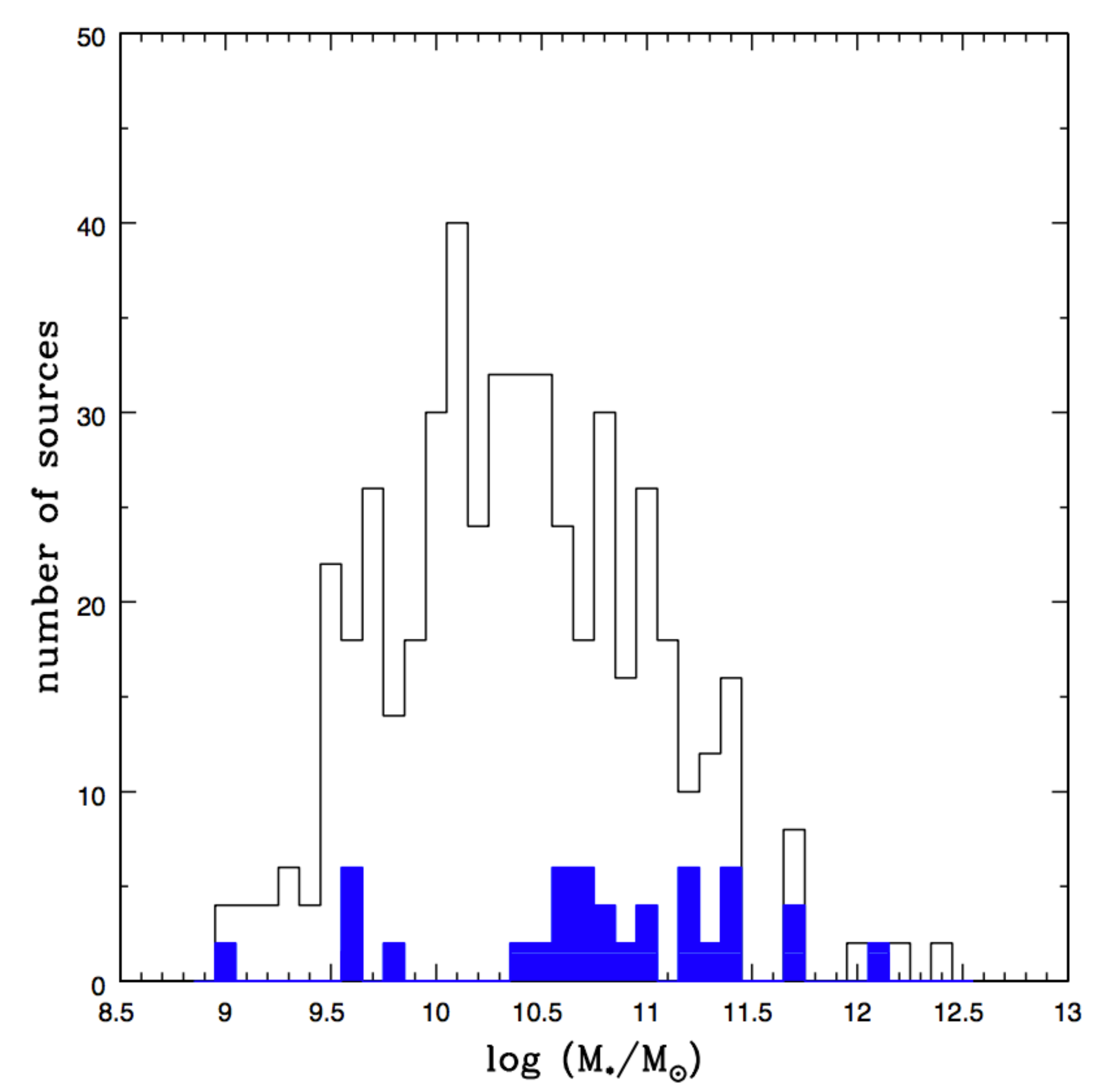}
  \label{fig_mstar_distrib_peculiar}
\end{subfigure}
\begin{subfigure}
  \centering
  \includegraphics[width=0.45\linewidth]{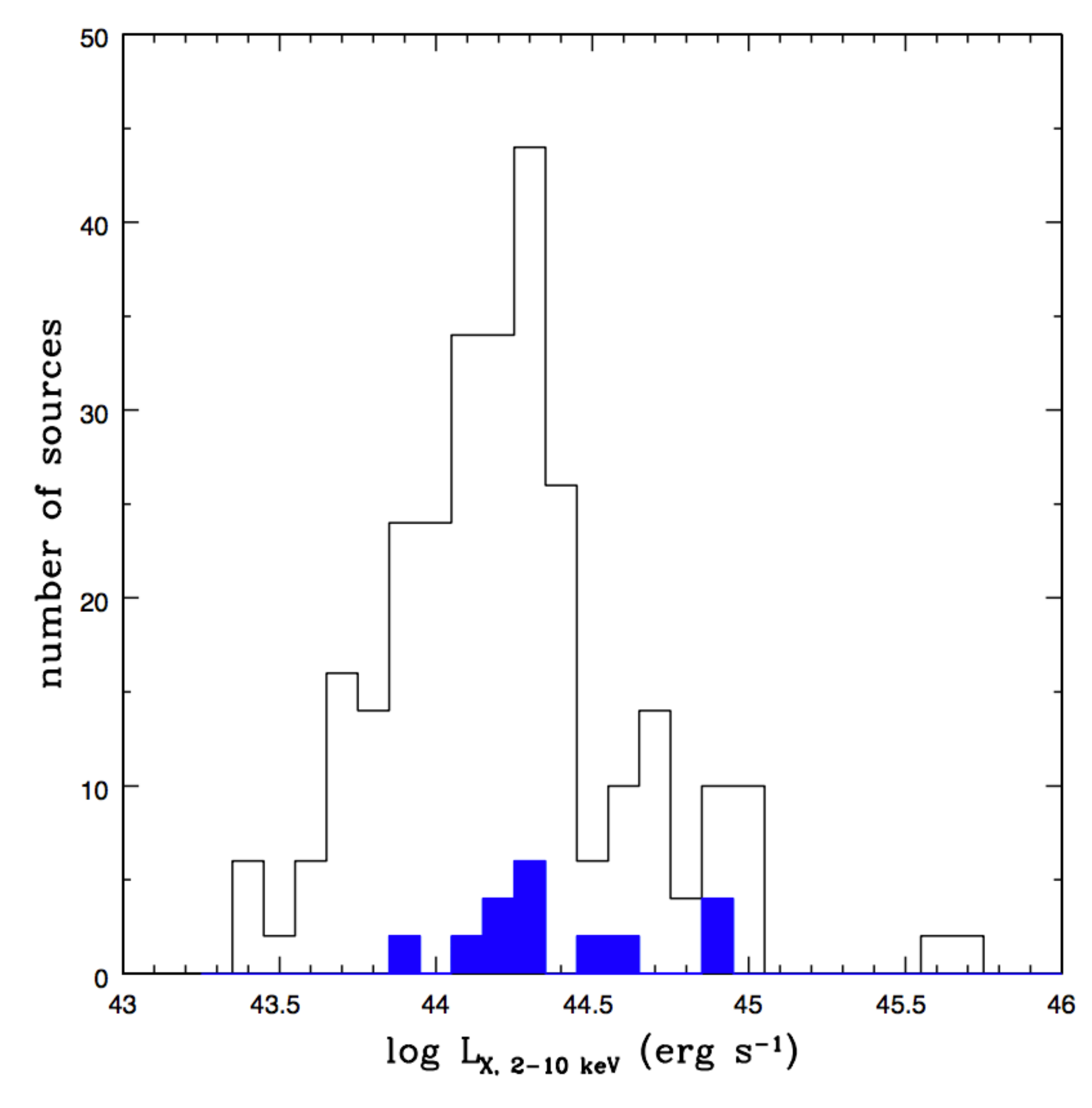}
  \label{fig_mstar_distrib_peculiar}
\end{subfigure}
\caption{Redshift, SFR, stellar mass calculations (derived by SED fitting) and X-ray luminosities (derived by X-ray spectral fitting), for the total population of X-ray unabsorbed, IR AGN (black line) and for the X-ray unabsorbed, optically red and obscured sources (blue shaded regions; see text for more details). We notice that the latter population consists of luminous AGN that (the bulk of it) lives in massive systems with $\log (\rm{M}_{\star} / \rm{M}_{\odot})>10.5$.}
\label{fig_peculiar}
\end{figure*}

In Fig. \ref{fig_peculiar} we plot the redshift, SFR, stellar mass (derived by SED fitting) and X-ray luminosity (derived by X-ray spectral fitting) distributions (blue shaded regions) of these 58 systems and compare them with the total population of X-ray unabsorbed, IR AGN. The plots indicate that these sources are X-ray luminous AGN that reside in massive systems ($\log (\rm{M}_{\star} / \rm{M}_{\odot})>10.5$).

\section{Summary}

We use the $\sim 300,000$ sources of the Stripe 82-XMM area to select mid-IR AGN by applying the selection criteria presented in \cite{Stern2012} combined with those described in \cite{Assef2013} for fainter sources \citep{Lamassa2019}. The sample consists of 1946 AGN candidates, with $90\%$ reliability. 
 For 1507 sources there are SDSS magnitudes available. Their optical/mid-IR colours $\rm r-W2>6$ \citep{Yan2013}, suggest that about 40\% of the sample is obscured. We use SED fitting to further investigate the optical obscuration of these sources, using the inclination angle as a proxy of the optical obscuration. The SED analysis corroborates that these sources are obscured.
 
507 IR AGN are detected in X-rays and have also optical spectra available. We examine the biases the aforementioned requirements impose on our final AGN sample. Our findings reveal that there is a (significant) population of red sources that are excluded when we require optical spectra and X-ray detection. Specifically, the percentage of red sources with SDSS detection drops from 43\% to 20\% when X-ray detection and optical spectra are required (Table \ref{table_yan}). Therefore, the sample used in our X-ray  analysis is biased against the redder sources. The X-ray spectral fitting analysis suggests that 8\% of the sources have column densities $\rm N_H>10^{22}~cm^{-2}$. There are 70 (15\% of the sample) optically red AGN, that are X-ray unasborbed. SEDs confirmed that 58/70 (83\%) of them are obscured sources. The remaining are galaxy dominated systems, i.e., the galaxy emission is a primary component of the SED in the optical part of the spectrum, even though the AGN emission is not absorbed.Therefore, our analysis shows that: (i) The IR acted AGN sample contains a significant fraction of optically red sources (ii) when we consider IR selected AGN that are also X-ray detected, our results reveal that these sources are preferentially associated with unobscured systems.

\begin{table*}
\caption{r-W2 values and spectral fitting results of the 58 sources, that although they are optically red and obscured based on their SEDs, they do not appear X-ray absorbed.}
\begin{tabular}{|l|l|l|l|l|l|l|l|l|l|l|l|}
\hline
  \multicolumn{1}{|c|}{RA} &
  \multicolumn{1}{c|}{DEC} &
  \multicolumn{1}{c|}{z} &
  \multicolumn{1}{c|}{r-W2} &
  \multicolumn{1}{c|}{Counts} &
  \multicolumn{1}{c|}{log ($\rm N_H/cm^{-2}$)}&
  \multicolumn{1}{c|}{$- \rm log\, \Delta N_H$} &
  \multicolumn{1}{c|}{$+\rm log \,\Delta N_H$} &
  \multicolumn{1}{c|}{$\Gamma$} &
  \multicolumn{1}{c|}{$-\Delta \Gamma$} &
  \multicolumn{1}{c|}{$+\Delta \Gamma$} &
  \multicolumn{1}{c|}{$\log \, L_X$} \\
   & & & & & &  & & & & & ($\rm 2-10\,keV$)  \\
\hline
  22.75191 & 0.28993 & 1.215 & 7.55 & 78 & 21.33 & 0.27 & 0.18 & 1.83 & 0.20 & 0.27 & 44.6\\
  14.17168 & -0.39824 & 0.732 & 7.22 & 42 & 21.69 & 0.24 & 0.18 & 1.88 & 0.22 & 0.26 & 43.7\\
  19.09421 & 0.12796 & 1.123 & 6.61 & 68 & 21.05 & 0.97 & 0.50 & 1.91 & 0.21 & 0.21 & 44.0\\
  15.91209 & 0.39683 & 1.018 & 6.59 & 57 & 21.83 & 1.69 & 0.28 & 1.86 & 0.24 & 0.23 & 44.2\\
  18.58634 & -0.39547 & 2.020 & 6.54 & 71 & 20.28 & 0.19 & 1.82 & 1.92 & 0.24 & 0.17 & 44.9\\
  24.64576 & 0.36487 & 0.799 & 6.39 & 21 & 20.65 & 0.59 & 0.85 & 1.91 & 0.25 & 0.23 & 44.2\\
  27.06873 & 0.23135 & 2.060 & 6.38 & 55 & 21.27 & 1.04 & 0.26 & 1.99 & 0.30 & 0.16 & 44.8\\
  18.53689 & 0.12753 & 0.748 & 6.36 & 36 & 21.25 & 1.14 & 0.37 & 1.83 & 0.20 & 0.29 & 44.1\\
  23.47359 & -0.25258 & 0.448 & 6.32 & 87 & 21.37 & 0.32 & 0.24 & 1.93 & 0.22 & 0.25 & 43.5\\
  19.70129 & -0.387586 & 1.498 & 6.26 & 42 & 21.61 & 0.92 & 0.49 & 1.80 & 0.21 & 0.27 & 44.0\\
  15.25358 & -0.39835 & 1.147 & 6.25 & 47 & 21.72 & 0.82 & 0.19 & 1.87 & 0.21 & 0.24 & 43.8\\
  14.15460 & -0.44718 & 0.474 & 6.24 & 252 & 20.12 & 0.10 & 0.64 & 2.02 & 0.18 & 0.15 & 44.3\\
  20.25309 & -0.25531 & 0.864 & 6.23 & 622 & 20.09 & 0.06 & 0.54 & 1.94 & 0.08 & 0.15 & 43.9\\
  25.77309 & 0.2756 & 1.048 & 6.20 & 252 & 20.18 & 0.13 & 1.07 & 1.85 & 0.12 & 0.19 & 44.6\\
  27.64483 & -0.03117 & 1.739 & 6.14 & 224 & 20.28 & 0.23 & 0.88 & 2.04 & 0.12 & 0.21 & 44.5\\
  23.88331 & 0.19228 & 1.152 & 6.11 & 68 & 20.43 & 0.36 & 1.12 & 1.98 & 0.23 & 0.23 & 44.3\\
  25.23941 & 0.1812 & 0.523 & 6.08 & 1970 & 20.00 & 0.02 & 0.05 & 1.99 & 0.07 & 0.07 & 44.5\\
  16.59543 & -0.35543 & 0.448 & 6.07 & 82 & 21.69 & 0.70 & 0.36 & 1.93 & 0.25 & 0.21 & 44.2\\
  26.54346 & -0.16881 & 1.016 & 6.06 & 68 & 20.11 & 0.03 & 1.42 & 1.93 & 0.19 & 0.22 & 44.3\\
  20.49188 & 1.13305 & 0.58 & 8.65	& 48	& 21.55	&1.42	& 0.31	&1.92	&0.18	& 0.29	&43.8   \\
  23.55290 & 0.06254 & 0.88 & 7.98	& 471 	&20.42	& 0.14	&0.17	&2.02	&0.12	&0.12	& 44.6\\
  26.38281 & -0.04591 & 1.03 & 7.88	& 65		&20.95	&0.62	&0.75	&1.66	&0.27	&0.25     &44.1\\
  22.75191 & 0.28993 & 1.19 & 7.55	& 276	&21.59	&0.20	&0.56	&1.94	&0.14	&0.15	&44.4\\
  25.02285 & -0.55147 & 0.75 & 7.26	& 776	&21.67	&0.05	&0.03	&1.88	&0.07	&0.13	&44.8\\
  16.94573 & -0.34984 & 0.57 & 7.04	& 56		&20.92	&0.84	&0.25	&2.12	&0.21	&0.28	&43.8\\
  18.86014 & -0.08849 & 0.88 & 6.89	&162		&20.33	&0.32	&0.40	&2.31	&0.14	&0.16	& 44.3\\
  27.35273 & -0.4=966 & 0.59 & 6.85	& 59		&20.75	&0.67	&0.46	&2.04	&0.19	&0.24	&43.6\\
  17.77068 & -0.27414 & 0.37 & 6.73	& 1309	&21.68	&0.02	&0.04	&2.06	&0.06	&0.14	&45.2\\
  22.71088 & -0.18812 & 1.35 & 6.70	& 87		&21.22	&1.14	&0.41	&1.92	&0.22	&0.31	&43.9\\
  19.09421 & 0.12796 & 1.12 & 6.61	& 150	&20.41	&0.37	&0.41	&1.92	&0.14	&0.18	&44.1\\
  13.66238 & -1.18152 & 1.37 & 6.59	& 51		&20.69	&0.42	&0.31	&2.07	&0.24	&0.23	&44.0\\
  26.98311 & 0.22810 & 1.64 & 6.54	& 138	&21.57	&0.40	&1.38	&2.24	&0.19	&0.15	&44.3\\
  18.58251 & -0.39484 & 2.02 & 6.54	& 92		&20.64	&0.48	&0.25	&2.01	&0.24	&0.25	&44.1\\
  21.37418 & -1.24827 & 0.82 & 6.48	& 29		&21.00	&0.61	&1.82	&1.81	&0.31	&0.29	&43.7\\
  15.14123 & 0.12564 & 1.06 & 6.45	& 73		&21.25	&0.45	&0.29	&1.75	&0.25	&0.24	&43.9\\
  24.63729 & 0.36478 & 0.80 & 6.39	& 484	&21.83	&0.20	&0.15	&2.07	&0.13	&0.14	&44.6\\
  18.53689 & 0.12753 & 0.75 & 6.36	& 133	&21.32	&0.29	&0.38	&1.85	&0.20	&0.19	&44.1\\
  26.82609 & -0.13351 & 0.54 & 6.36	& 123	&21.58	&0.45	&0.29	&2.05	&0.18	&0.15	&44.4\\
  16.39569 & 0.10205 & 1.30 & 6.32	& 132	&20.86	&0.42	&1.21	&1.75	&0.21	&0.16	&44.3\\
  27.78665 & 0.58813 & 0.88 & 6.31	& 55		&20.81	&0.65	&0.34	&2.10	&0.26	&0.21	&43.9\\
  19.70257 & -0.38753 & 1.50 & 6.26	& 76		&20.50	&0.40	&0.82	&1.62	&0.17	&0.28	&43.9\\
  26.52557 & 0.32857 & 0.90 & 6.25	&1083	&21.69	&0.04	&0.02	&2.15	&0.05	&0.12	&44.6\\
  14.15460 & -0.44718 & 0.47 & 6.24	&111		&20.24	&0.24	&0.82	&1.94	&0.24	&0.21	&44.2\\
  16.35091 & 0.15034 & 0.42 & 6.24	& 69		&20.69	&0.62	&0.45	&1.85	&0.28	&0.29	&44.0\\
  20.25309 & -0.25531 & 0.86 & 6.23	&100		&21.24	&0.47	&0.28	&1.92	&0.19	&0.25	&44.2\\
  18.81533 & 0.4096 & 1.94 & 6.19	& 73		&21.42	&0.96	&0.61	&1.97	&0.21	&0.18	&43.9\\
  16.23248 & 0.28018 & 0.72 & 6.13	&531		&21.69	&0.05	&0.8		&2.35	&0.09	&0.14	&44.6\\
  22.89568 & 0.18808 & 0.90 & 6.12	& 92		&21.52	&0.43	&0.21	&2.01	&0.26	&0.25	&44.1\\
  18.30137 & 0.10406 & 0.97 & 6.11	& 127	&21.50	&0.31	&0.39	&1.75	&0.17	&0.19	&44.1\\
  16.47210 & -0.01478 & 1.05 & 6.09	& 185	&21.00	&0.52	&0.74	&1.89	&0.19	&0.14	&44.2\\
  26.88057 & -0.15136 & 2.24 & 6.09	& 127	&21.82	&0.32	&0.24	&1.92	&0.16	&0.18	&44.0\\
  28.26243 & 1.01048 & 1.31 & 6.08	& 57		&20.91	&0.52	&1.91	&1.95	&0.27	&0.21	&43.8\\
  16.59543 & -0.35543 & 0.45 & 6.07	&112		&20.39	&0.32	&0.25	&2.08	&0.21	&0.20	&44.0\\
  19.61204 & -0.49189 & 1.16 & 6.07	& 71		&21.52	&1.21	&0.58	&1.72	&0.22	&0.29	&43.9\\
  27.80391 & 0.50809 & 0.70 & 6.04	&204		&20.20	&0.25	&1.21	&1.92	&0.15	&0.15	&44.4\\
  26.88310 & -0.33868 & 0.55 & 6.04	& 50		&21.05	&0.49	&0.73	&2.14	&0.25	&0.27	&43.8\\
  27.27786 & -0.37372 & 1.07 & 6.01	&225		&21.30	&0.18	&0.27	&1.83	&0.16	&0.14	&44.5\\
  28.77457 & 0.59739 & 1.07 & 6.00	&168		&21.69	&0.34	&0.51	&1.99	&0.14	&0.18	&44.3\\
 
\hline
\end{tabular}
\label{table_58_sources}
\end{table*}

\section{Acknowledgments}

The authors are grateful to the anonymous referee for helpful comments. GM and AR acknowledge support of this work by the PROTEAS II project (MIS 5002515), which is implemented under the "Reinforcement of the Research and Innovation Infrastructure" action, funded by the "Competitiveness, Entrepreneurship and Innovation" operational programme (NSRF 2014-2020) and co-financed by Greece and the European Union (European Regional Development Fund).

This research has made use of data obtained from the 3XMM XMM-\textit{Newton} 
serendipitous source catalogue compiled by the 10 institutes of the XMM-\textit{Newton} 
Survey Science Centre selected by ESA.

This work is based on observations made with XMM-\textit{Newton}, an ESA science 
mission with instruments and contributions directly funded by ESA Member States 
and NASA. 

Funding for the Sloan Digital Sky Survey IV has been provided by the Alfred P. Sloan Foundation, the U.S. Department of Energy Office of Science, and the Participating Institutions. SDSS-IV acknowledges
support and resources from the Center for High-Performance Computing at
the University of Utah. The SDSS web site is www.sdss.org.

SDSS-IV is managed by the Astrophysical Research Consortium for the 
Participating Institutions of the SDSS Collaboration including the 
Brazilian Participation Group, the Carnegie Institution for Science, 
Carnegie Mellon University, the Chilean Participation Group, the French Participation Group, Harvard-Smithsonian Center for Astrophysics, 
Instituto de Astrof\'isica de Canarias, The Johns Hopkins University, 
Kavli Institute for the Physics and Mathematics of the Universe (IPMU) / 
University of Tokyo, Lawrence Berkeley National Laboratory, 
Leibniz Institut f\"ur Astrophysik Potsdam (AIP),  
Max-Planck-Institut f\"ur Astronomie (MPIA Heidelberg), 
Max-Planck-Institut f\"ur Astrophysik (MPA Garching), 
Max-Planck-Institut f\"ur Extraterrestrische Physik (MPE), 
National Astronomical Observatories of China, New Mexico State University, 
New York University, University of Notre Dame, 
Observat\'ario Nacional / MCTI, The Ohio State University, 
Pennsylvania State University, Shanghai Astronomical Observatory, 
United Kingdom Participation Group,
Universidad Nacional Aut\'onoma de M\'exico, University of Arizona, 
University of Colorado Boulder, University of Oxford, University of Portsmouth, 
University of Utah, University of Virginia, University of Washington, University of Wisconsin, 
Vanderbilt University, and Yale University.

\bibliography{mybib}{}
\bibliographystyle{mn2e}

\appendix

\section{SED fitting}
\label{sec_sed_fitting}

We construct SEDs for the 507 IR AGN in our sample, with X-ray detection. For that purpose, we use multiwavelength broadband photometric data from optical \citep[SDSS-DR13;][]{Albareti2015}, near-IR \citep[VISTA-VIKING;][]{Emerson2006, Dalton2006} and mid-IR \citep[WISE;][]{Wright2010}. The SEDs are fitted using the CIGALE code version 0.12 \citep[Code Investigating GALaxy Emission;][]{Noll2009}. 

In the fitting process we use the \cite{Fritz2006} library of templates to model the AGN emission. The star-formation histories are convolved assuming the double-exponentially-decreasing (2$\tau$-dec) model \citep{Ciesla2015}. The \cite{Bruzual_Charlot2003} template is utilized to model the stellar population systhesis, assuming the Salpeter Initial Mass Function (IMF) and the \cite{Calzetti2000} dust extinction law. \cite{Dale2014} templates are adopted for the absorbed dust that is reemitted in the IR. Table \ref{table_cigale} presents the models and the values for the free parameters used by CIGALE for the SED fitting of our X-ray AGN. We consider a source as type 2 when the inclination angle of the torus, $\Psi$, is $\Psi=0.001^{o}$ or $\Psi=50.10^{o}$.

\begin{table*}
\caption{The models and the values for their free parameters used by CIGALE for the SED fitting of our X-ray AGN. $\tau$ is the e-folding time of the main stellar population model in Myr, age is the age of the main stellar population in the galaxy in Myr (the precision is 1\,Myr), burst age is the age of the late burst in Myr (the precision is 1 Myr). $\beta$ and $\gamma$ are the parameters used to define the law for the spatial behaviour of density of the torus density. The functional form of the latter is $\rho (r, \theta) \propto  r^\beta e^{-\gamma | cos \theta |}$, where r and $\theta$ are the radial distance and the polar distance, respectively. $\Theta$ is the opening angle and $\Psi$ the inclination angle of the torus. Type-2 AGN have $\Psi=0.001^{o}$ and Type-1 AGN have $\Psi=89.99^{o}$. The AGN fraction is measured as the AGN emission relative to IR luminosity ($1-1000\,\mu m$).} 
\centering
\setlength{\tabcolsep}{0.1mm}
\begin{tabular}{cc}
       \hline
Parameter &  Model/values \\
	\hline
\multicolumn{2}{c}{Stellar population synthesis model} \\
\\
Initial Mass Function & Salpeter\\
Metallicity & 0.02 (Solar) \\
Single Stellar Population Library & Bruzual \& Charlot (2003) \\
\hline
\multicolumn{2}{c}{double-exponentially-decreasing (2$\tau$-dec) model} \\
$\tau$ & 100, 1000, 5000, 10000 \\ 
age & 500, 2000, 5000, 10000, 12000 \\
burst age & 100, 200, 400 \\
\hline
\multicolumn{2}{c}{Dust extinction} \\
Dust attenuation law & \cite{Calzetti2000} \\
Reddening E(B-V) & 0.3 \\ 
E(B-V) reduction factor between old and young stellar population & 0.44 \\
\hline
\multicolumn{2}{c}{\cite{Fritz2006} model for AGN emission} \\ 
Ratio between outer and inner dust torus radii &    60 \\
9.7 $\mu m$ equatorial optical depth & 0.1, 0.3, 1.0, 2.0, 6.0, 10.0 \\
$\beta$ &-1.00, -0.5, 0.00 \\
$\gamma$ & 0.0, 2.0, 6.0 \\
$\Theta$ & 100 \\
$\Psi$ & 0.001, 50.10, 89.99 \\
AGN fraction & 0.1, 0.2, 0.3, 0.5, 0.6, 0.8 \\
\hline
\label{table_cigale}
\end{tabular}
\end{table*}

\end{document}




\maketitle

\label{firstpage}

\begin{abstract}
TBD
\end{abstract}

\begin{keywords}
galaxies: active, galaxies: haloes, galaxies: Seyfert, X-rays: diffuse
background 
\end{keywords}

\section{Introduction}

\appendix

\section{MUSYC}

In  the ECDFS  we use  the MUSYC  Subaru v1.0  catalog  which provides
photometry in 32  bands from the UV to  the mid-infrared (Cardamone et
al. 2009). For  the DEEP2 we use the photometric  catalogue of Coil et
al. (2004).  As  there is no overlap between the MUSYC  and any of the
DEEP2 fields, the colour  transformation between the filtersets in the
two  photometric  surveys are  determined  using theroretical  stellar
tracks. The LePHARE  code (Arnouts et al. 1999; Ilbert  et al 2006) is
used to convolve the DEEP2 and MUSYC $B$, $R$ and $I$ filters with the
Pickles  (1998) stellar  templates  and predict  the  $B-R$ and  $R-I$
colours    of   stars    in   the    two   filtersets.     In   Figure
\ref{fig_musyc_deep2_stars} these stellar tracks are compared with the
photometry of optically unresolved sources (i.e. stellar like) in each
of  the two  surveys.  In  the DEEP2  catalogue these  are  defined as
objects with  $P_{GAL}<0.2$ (probability that the  source is resolved,
see Coil  et al.  2004  for details) and $19<R_{\rm  D2}<23$\,mag.  In
the case of  the MUSYC catalogue stars are  selected by requiring that
the parameter {\sc star\_flag} equals unity (see Cardamone et al. 2009
for      details)     and     $19<R_{\rm      M}<23$\,mag.      Figure
\ref{fig_musyc_deep2_stars} shows that  the photometric calibration of
the MUSYC  and DEEP2  surveys is offset  from the  theoretical stellar
tracks  by about $+0.05$  and $+0.15$\,mag  respectively in  the $R-I$
colour.  These  systematic offsets should  be taken into  account when
converting from the MUSYC to  the DEEP2 filterset. The source of these
offsets is beyond the scope of the analysis presented here.

Next we compare  the MUSYC and DEEP2 filtersets  by plotting in Figure
\ref{fig_musyc_deep2} the $(B-R)_{\rm  M}$ versus the $(B-R)_{\rm D2}$
and the  $(R-I)_{\rm M}$ against  the $(R-I)_{\rm D2}$ colours  of the
Pickles (1998) stellar templates. In the notation above the subscripts
$\rm M$  and $\rm D2$ denote  the MUSYC and  DEEP2 bands respectively.
For  simplicity, linear  relations are  fit to  the stellar  tracks in
Figure  \ref{fig_musyc_deep2}  to transform  the  $(B-R)_{\rm M}$  and
$(R-I)_{\rm  M}$  colours to  $(B-R)_{\rm  D2}$  and $(R-I)_{\rm  D2}$
respectively.   We   have  also  experimented   fitting  higher  order
polynomials  to  the data  points  in Figure\ref{fig_musyc_deep2}  but
found  the  final  results  and  conclusions  remain  unchanged.   The
best-fit     linear    relations     are     presented    in     Table
\ref{tab_musyc_deep2}.

Figure \ref{fig_musyc_deep2_rband} plots $R_{\rm M}-R_{\rm D2}$ against
$(R-I)_{\rm M}$  for the Pickles  (1998) stellar templates.   It shows
that $R_{\rm  M}-R_{\rm D2}$ is  well correlated with  the $(R-I)_{\rm
M}$  colour up  to $(R-I)_{\rm  M}\approx0.5$. At  redder  colours the
relation appears to flatten and the scatter in the theoretical stellar
locus increases.  We choose to fit a linear relation to the datapoints
up to  $(R-I)_{\rm M}=0.5$.  At redder colours,  the $R_{\rm M}-R_{\rm
D2}$ is approximated with a constant  fixed to the value of the linear
relation at $(R-I)_{\rm M}=0.5$.  Although this approximation may lead
to  systematic uncertainties  in the  conversion from  $R_{\rm  M}$ to
$R_{\rm D2}$, these are expected to be at the level of few tenths of a
magnitude.  Table  \ref{tab_musyc_deep2} presents the  functional form
used to  describe the  variations of the  $R_{\rm M}-R_{\rm  D2}$ with
$(R-I)_{\rm M}$.

As a consistency  check of the colour transformations  listed in Table
\ref{tab_musyc_deep2}   we   compare    the   magnitude   and   colour
distributions of  galaxies in the  DEEP2 and MUSYC surveys.   For each
source in the MUSYC  photometric catalogue we estimate its $(B-R)_{\rm
D2}$,  $(R-I)_{D2}$ colours  and  $R_{D2}$ magnitude  by applying  the
linear relations of Table  \ref{tab_musyc_deep2}. The number counts in
the $R_{D2}$ band are then constructed for MUSYC sources classified as
galaxies ({\sc star\_flag} parameter equals null; see Cardamone et al.
2009  for details).  These  are then  compared with  the $R_{D2}$-band
number counts of DEEP2 galaxies (extended source probability parameter
$P_{GAL}>0.2$; see Coil et al.   2004 for details).  The galaxy counts
normalised  to the  area of  each  photometric survey  are plotted  in
Figure \ref{fig_musyc_deep2_counts}.  There  is a systematic offset of
about  0.4\,mag  between  the  two  photometric  catalogues.   Similar
offsets are also apparent in  the $B_{D2}$ and $I_{D2}$ galaxy counts.
This suggests  that the discrepancy  is because of differences  in the
determination  of total  magnitudes  in  the two  surveys  and is  not
related to the colour  transformation between the two filtersets. This
is  further supported  by  Figure \ref{fig_musyc_deep2_colours}  which
compares the  $(B-R)_{\rm D2}$  and $(R-I)_{D2}$ distributions  of the
MUSYC and  DEEP2 surveys. These are constructed  by selecting galaxies
in the two surveys with $18<R_{D2}<24$\,mag. The faint magnitude limit
is chosen  to minimise the  effect of incompleteness in  the shallower
DEEP2 photometric survey. In the case of the MUSYC catalogue an offset
of +0.4\,mag is  applied to the estimated $R_{D2}$  to account for the
systematic offset  in the  number counts of  the two surveys.   In the
case of  the $(R-I)_{D2}$ colour distribution the  offsets between the
the  theoretical stellar  tracks and  the MUSYC/DEEP2  photometry (see
Fig.   \ref{fig_musyc_deep2_stars})  have  also  been applied  to  the
data. The  agreement between the MUSYC and  DEEP2 colour distributions
in  Figure  \ref{fig_musyc_deep2_colours}  indicates that  the  colour
transformations of  Table \ref{tab_musyc_deep2} are robust  and can be
used to  convert the fluxes  measured throught the MUSYC  filterset to
the DEEP2 $BRI$ bands.

\begin{figure*}
\begin{center}
\includegraphics[height=0.9\columnwidth]{musyc_BR_RI_stars.pdf}
\includegraphics[height=0.9\columnwidth]{deep2_BR_RI_stars.pdf}
\end{center}
\caption{$B-R$ vs  $R-I$ stellar locus. The  panel of the  left is for
the MUSYC filterset (subscript ``M'' in $B-R$ and $R-I$ colours).  The
right panel is for the DEEP2 bands (subscript ``D2'').  In both panels
the black dots are optically  unresolved sources detected in the MUSYC
(left) and  DEEP2 (right) photometric surveys.  The  red triangles are
the theoretical  stellar tracks  determined by convolving  the Pickles
(1998) stellar template library with  the MUSYC (left panel) and DEEP2
(right panel) $BRI$ filters.  }\label{fig_musyc_deep2_stars}
\end{figure*}

\begin{figure*}
\begin{center}
\includegraphics[height=0.9\columnwidth]{musyc_BRm_BRd2.pdf}
\includegraphics[height=0.9\columnwidth]{musyc_RIm_RId2.pdf}
\end{center}
\caption{{\bf  Left panel:} $(B-R)_{\rm  M}$ colour  estimated through
the MUSYC bands against $(B-R)_{\rm D2}$ colour determined through the
DEEP2  filters.  The  red  triangles are  the  Pickles (1998)  stellar
templates. The  black line shows  the best-fit linear relation  to the
theoretical stellar  track. {\bf Right panel:}  $(R-I)_{\rm M}$ colour
estimated  through the  MUSYC  bands against  $(R-I)_{\rm D2}$  colour
determined  through the  DEEP2  filters.  The  red  triangles are  the
Pickles (1998)  stellar templates. The  black line shows  the best-fit
linear     relation    to     the    theoretical     stellar    track.
}\label{fig_musyc_deep2}
\end{figure*}

\begin{figure}
\begin{center}
\includegraphics[height=0.9\columnwidth]{musyc_RmRd2_RIm.pdf}
\end{center}
\caption{$R_{\rm  M}-R_{\rm D2}$ against  $(R-I)_{\rm M}$  colour. The
red triangles are the Pickles (1998) stellar templates. The black line
shows the linear relation  used to approximate the theoretical stellar
track.  }\label{fig_musyc_deep2_rband}
\end{figure}

\begin{figure}
\begin{center}
\includegraphics[height=0.9\columnwidth]{musyc_hist_Rd2.pdf}
\end{center}
\caption{Galaxy number  counts in 0.1\,mag  bins for the  $R_{\rm D2}$
filter.   The  red  filled  circles  are  for  the  MUSYC  photometric
catalogue.  The  blue line  is for the  DEEP2 photometric  survey. The
number  counts  are  normalised  to  the area  of  each  survey,  $\rm
0.33\,deg^2$ in the case of  MUSYC and $\rm 1.18\,deg^2$ for the DEEP2
survey  of the  Extended Groth  Strip. There  is systematic  offset of
about 0.4\,mag between the two surveys.
}\label{fig_musyc_deep2_counts}
\end{figure}

\begin{figure}
\begin{center}
\includegraphics[height=0.9\columnwidth]{hist_colour_br.pdf}
\includegraphics[height=0.9\columnwidth]{hist_colour_ri.pdf}
\end{center}
\caption{$(B-R)_{\rm  D2}$ and  $(R-I)_{D2}$  colour distributions  in
0.1\,mag bins of galaxies selected in the MUSYC and DEEP2 surveys. For
MUSYC   sources   the  colours   are   estimated   using  the   colour
transformations   of    Table   \ref{tab_musyc_deep2}.    The   colour
distribution  is   limited  to  galaxies  in  the   two  surveys  with
$18<R_{D2}<24$\,mag.  }\label{fig_musyc_deep2_colours}
\end{figure}

\section{CFHTLS}

The CFHTLS  D3 field  overlaps with the  DEEP2 survey in  the Extended
Groth Strip. It is  therefore possible to determine the tranformations
from  the CFHTLS  $ugriz$ bands  to  the DEEP2  $BRI$ filterset  using
sources that are common in the two surveys.  The data release T0004 of
the   CFHTLS  photometric   catalogue   (Coupon  et   al.   2009)   is
cross-matched with the  DEEP2 source list (Coil et  al.  2004) using a
search  radius of  2\,arcsec. Regions  of poor  photometry  because of
nearby  bright stars  or  difraction  spikes are  masked  out in  this
exercise.  

The transformations of the $g-r$ and $r-i$ colours to $(B-R)_{\rm D2}$
and  $(R-I)_{\rm  D2}$  respectively  are determined  using  optically
resolved  sources  in  the  deeper CFHTLS  catalogue  (parameter  {\sc
flag_terapix}    equals   null,    Coupon   et    al.     2009)   with
$18<R_{D2}<24$\,mag, i.e.  in the magnitude range of galaxies followed
with DEEP2 spectroscopy.   We choose to use galaxies  to determine the
transformation between colours in the  two filtersets to have a handle
on  the scatter  expected  in those  relation  because of  photometric
uncertainties, the different approaches for determining colours in the
two  surveys  as  well  as   the  diversity  of  the  Spectral  Energy
Distributions       of       extragalactic      sources.        Figure
\ref{fig_cfhtls_deep2}   compares   the   CFHTLS   and   DEEP2
filtersets by plotting  for galaxies in the overlap  region of the two
surveys the  $(B-R)_{\rm D2}$ colour versus $g-r$  and the $(R-I)_{\rm
D2}$  colour against  $r-i$.  The  mean $(B-R)_{\rm  D2}$ [$(R-I)_{\rm
D2}$] and its  standard deviation within $g-r$ ($r-i$)  colour bins of
0.1\,mag size are also shown  in Figure XX.  A second order polynomial
is fit to  the binned datapoints in the  $(B-R)_{\rm D2}$ versus $g-r$
plot,  while  a  linear  relation   is  used  for  the  $r-i$  against
$(R-I)_{\rm D2}$ colour plot.  The best-fit coefficients are listed in
Table \ref{tab_cfhtls_deep2}.

\begin{figure*}
\begin{center}
\includegraphics[height=0.9\columnwidth]{colour_gr_BR.pdf}
\includegraphics[height=0.9\columnwidth]{colour_ri_RI.pdf}
\end{center}
\caption{{\bf   Left:}   $(B-R)_{\rm   D2}$  plotted   against   $g-r$
colour. The  black dots are resolved  sources in the D3  region of the
CFHTLS (parameter {\sc flag_terapix} equals null, Coupon et al.  2009)
with DEEP2 photometric survey counterparts and magnitudes in the range
$18<R_{D2}<24$\,mag. The mean $(B-R)_{\rm D2}$ and 1\,sigma rms within
$g-r$  bins  (0.1\,mag  size)  are  shown with  the  red  circles  and
errorsbars.  The  red line  is the best-fit  linear relation  to those
datapoints. The rms scatter around  this line is estimated to be about
0.15\,mag.   {\bf  Right:}  $(R-I)_{\rm  D2}$  plotted  against  $r-i$
colour. The  black dots are resolved  sources in the D3  region of the
CFHTLS    with   DEEP2   counterparts    in   the    magnitude   range
$18<R_{D2}<24$\,mag. The mean $(R-I)_{\rm D2}$ and 1\,sigma rms within
$r-i$  bins  (0.1\,mag  size)  are  shown with  the  red  circles  and
errorsbars.  The red curve is  the best-fit second order polynomial to
those datapoints. The rms scatter  around this line is estimated to be
about 0.1\,mag.  
}\label{fig_cfhtls_deep2}
\end{figure*}

\begin{figure}
\begin{center}
\includegraphics[height=0.9\columnwidth]{colour_ri_Ri.pdf}
\end{center}
\caption{$R_{\rm D2}-r$  plotted against $r-i$ colour.  The black dots
are unresolved sources in the  D3 region of the CFHTLS (parameter {\sc
flag_terapix}  equals   unity,  Coupon   et  al.   2009)   with  DEEP2
counterparts  in  the magnitude  range  $18<R_{D2}<24$\,mag. The  mean
$R_{\rm D2}-r$ and 1\,sigma rms  within $r-i$ bins (0.2\,mag size) are
shown with the  red circles and errorsbars.  The red  line is the best
linear fit to those datapoints.  }\label{fig_cfhtls_deep2_Rr}
\end{figure}

\begin{figure}
\begin{center}
\includegraphics[height=0.9\columnwidth]{fig_cfhtls_deep2_Rr.pdf}
\end{center}
\caption{Galaxy  number counts in  $R_{\rm D2}$  band. The  red curves
correspond to  the 1\,sigma rms envelope  of the galaxy  counts in the
DEEP2  photometric  survey  of  the EGS.   These  observations  become
incomplete  fainter than $R\approx24$\,mag  resulting in  the observed
turnover at  faint magnitudes.  The blue dots  are the  infered galaxy
number counts  in the  CFHTLS $r$-band converted  to the  $R_{\rm D2}$
filter using the transformations listed in Table \ref{tab_cfhtls_deep2}.
}\label{fig_cfhtls_deep2_counts}
\end{figure}

The conversion from the CFHTLS $r$-band to the DEEP2 $R_{\rm D2}$ uses
optically unresolved  sources (i.e.  stars).  This is  because in this
case one needs objects with accurate photometry that is least affected
by  systematics related  to  e.g.  the  extent  of the  source or  the
determination of aperture  corrections. We select optically unresolved
sources  in the  CFHTLS  (parameter {\sc  flag_terapix} equals  unity,
Coupon et  al.  2009)  with DEEP2 counterparts  and magnitudes  in the
range  $18<R_{D2}<23$\,mag. We choose  not to  use fainter  sources to
keep  photometric errors  small  and also  to  avoid contamination  by
galaxies at fainter magnitudes. Figure \ref{fig_cfhtls_deep2_Rr} plots
$R_{\rm D2}-r$ against $r-i$. Also  shown are the average and standard
deviation of  $R_{\rm D2}-r$ within  $r-i$ colour bins of  0.2\,mag in
size.  The best-fit  linear relation to the data  is plotted in Figure
\ref{fig_cfhtls_deep2_Rr}     and     is     presented    in     Table
\ref{tab_cfhtls_deep2}. The  $r$-band manitude  of each galaxy  in the
CFHTLS D3  field is  converted to $R_{\rm  D2}$ and  the corresponding
galaxy number counts  in that filter are constructed.   These are then
compared  with the galaxy  number counts  estimated directly  from the
DEEP2   photometric    survey   of    the   EGS   field    in   Figure
\ref{fig_cfhtls_deep2_counts}.   There is  no evidence  for systematic
offsets in the two distributions.

We  also investigate  the accuracy  of the  transformations  listed in
Table  \ref{tab_cfhtls_deep2} by  comparing the  $(B-R)_{\rm  D2}$ and
$(R-I)_{\rm  D2}$ distributions  of galaxies  infered from  the CFHTLS
$gri$  photometry   with  those  estimate  directly   from  the  DEEP2
photometric survey  in the EGS.  In  this excersise we  also take into
account  the scatter of  datapoints around  the best-fit  relations in
Figure  \ref{fig_cfhtls_deep2}.    The  rms  scatter   of  the
$(B-R)_{\rm   D2}$  versus   $g-r$   relation  is   estimated  to   be
0.15\,mag. For the $(R-I)_{\rm D2}$  versus $r-i$ plot we determine an
rms scatter  of 0.1\,dex.   These numbers include  photometric errors,
differences in the determination of colours in the two surveys as well
as any intrinsic  scatter in those relations associated  with e.g. the
diversity  of the galaxy  SEDs. For  each CFHTLS  galaxy we  apply the
conversions  listed  in  Table  \ref{tab_cfhtls_deep2}  to  infer  its
$(B-R)_{\rm D2}$ and $(R-I)_{\rm D2}$. We then added Gaussian deviates
to those  colours with  Half Width Half  Maximum of 0.15  and 0.1\,dex
respectively.   The resulting colour  distributions are  compared with
those  determined for  galaxies  in the  DEEP2  photometric survey  in
Figure  \ref{fig_cfhtls_deep2_colours}.  This comparison  is only
for resolved sources in  the two surveys with $18<R_{\rm D2}<24$\,mag.
There  is   fair  agreement  between  the  CFHTLS   and  DEEP2  colour
distributions suggesting that at  least to the first approximation the
colour transformations of  Table \ref{tab_cfhtls_deep2} are robust and
can  be  used to  convert  the  fluxes  measured throught  the  CFHTLS
filterset to the DEEP2 $BRI$ bands.

\begin{figure*}
\begin{center}
\includegraphics[height=0.9\columnwidth]{hist_colour_BR.pdf}
\includegraphics[height=0.9\columnwidth]{hist_colour_RI.pdf}
\end{center}
\caption{$(B-R)_{\rm  D2}$ and  $(R-I)_{D2}$  colour distributions  in
0.1\,mag bins of galaxies selected in the CFHTLS-D3 and EGS DEEP2
surveys. For  CFHTLS-D3  sources   the  colours   are   estimated
using  the   colour 
transformations   of    Table   \ref{tab_cfhtls_deep2}.    The   colour
distribution  is   limited  to  galaxies  in  the   two  surveys  with
$18<R_{D2}<24$\,mag.  }\label{fig_cfhtls_deep2_colours}
\end{figure*}